\documentclass[fleqn,usenatbib,useAMS]{mnras}

\usepackage{graphicx}	
\usepackage{amsmath}	
\usepackage{amssymb}	
\usepackage{multicol}        
\usepackage{bm}		
\usepackage{pdflscape}	
\usepackage{verbatim}
\usepackage{mn2e-breakabs}
\usepackage{soul}

\def\xirppi{\xi^s(r_p,\pi)}
\def\({\left(}
\def\){\right)}
\def\[{\left[}
\def\]{\right]}
\def\mhkpc{\,h^{-1}{\rm kpc}}
\def\mhmpc{\,h^{-1}{\rm Mpc}}

\def\mhmpcc{\,h^{-3}{\rm Mpc^3}}
\def\mhgpcc{\,h^{-3}{\rm Gpc^3}}

\def\mwp{w_p(r_p)}

\newcommand*\wdd{w_{\rm{ang}}^{\rm{DD}}(\theta)}
\newcommand*\wdr{w_{\rm{ang}}^{\rm{DR}}(\theta)}

\DeclareMathOperator{\popcnt}{popcnt}

\usepackage[T1]{fontenc}
\usepackage{ae,aecompl}

\usepackage{newtxtext,newtxmath}

\title[DR16 eBOSS PIP+ANG Corrections for Fibre Collisions]{The Completed SDSS-IV extended Baryon Oscillation Spectroscopic Survey: Pairwise-Inverse-Probability and Angular Correction for Fibre Collisions in Clustering Measurements}

\author[F.  G.  Mohammad et al.]{\parbox{\textwidth}{
Faizan G. Mohammad$^{1,2}$\thanks{Email: faizan.mohammad@uwaterloo.ca},
Will J. Percival$^{1,2,3}$,
Hee-Jong Seo$^{4}$,
Michael J. Chapman$^{1,2}$,
D. Bianchi$^{5}$,
Ashley J. Ross$^{6}$,
Cheng Zhao$^{7}$,
Dustin Lang$^{3}$,
Julian Bautista$^{8}$,
Jonathan Brinkmann$^{9}$,
Joel R. Brownstein$^{10}$,
Etienne Burtin$^{11}$,
Chia-Hsun Chuang$^{12}$,
Kyle S. Dawson$^{10}$,
Sylvain de la Torre$^{13}$,
Arnaud de Mattia$^{11}$,
Sarah  Eftekharzadeh$^{10}$,
Sebastien Fromenteau$^{14}$,
H\'ector Gil-Mar\'in$^{5,15}$,
Jiamin Hou$^{16}$,
Eva-Maria Mueller$^{17,8}$,
Richard Neveux$^{11}$,
Romain Paviot$^{13}$,
Anand Raichoor$^{7}$,
Graziano Rossi$^{18}$,
Donald P. Schneider$^{19,20}$,
Am\'elie Tamone$^{7}$,
Jeremy L. Tinker$^{21}$,
Rita Tojeiro$^{22}$,
Mariana Vargas Maga\~na$^{23}$,
Gong-Bo Zhao$^{24,25}$
 } \vspace*{4pt} \\ 
\scriptsize $^{1}$ Waterloo Centre for Astrophysics, Department of Physics and Astronomy, University of Waterloo, Waterloo, ON N2L 3G1, Canada\vspace*{-2pt} \\ 
\scriptsize $^{2}$ Department of Physics and Astronomy, University of Waterloo, Waterloo, ON N2L 3G1, Canada\vspace*{-2pt} \\ 
\scriptsize $^{3}$ Perimeter Institute for Theoretical Physics, 31 Caroline St. North, Waterloo, ON N2L 2Y5, Canada\vspace*{-2pt} \\ 
\scriptsize $^{4}$ Department of Physics and Astronomy, Ohio University, 251B Clippinger Labs, Athens, OH 45701, USA\vspace*{-2pt} \\ 
\scriptsize $^{5}$ Institut de Ci\`encies del Cosmos, Universitat de Barcelona, ICCUB, Mart\'i i Franqu\`es 1, E08028 Barcelona, Spain\vspace*{-2pt} \\ 
\scriptsize $^{6}$ Center for Cosmology and Astro-Particle Physics, Ohio State University, Columbus, Ohio, USA\vspace*{-2pt} \\ 
\scriptsize $^{7}$ Institute of Physics, Laboratory of Astrophysics, \'Ecole Polytechnique F\'ed\'erale de Lausanne (EPFL), Observatoire de Sauverny, 1290 Versoix, Switzerland\vspace*{-2pt} \\ 
\scriptsize $^{8}$ Institute of Cosmology \& Gravitation, Dennis Sciama Building, University of Portsmouth, Portsmouth, PO1 3FX, UK\vspace*{-2pt} \\ 
\scriptsize $^{9}$ Apache Point Observatory and New Mexico State University, P.O. Box 59, Sunspot, NM 88349, USA\vspace*{-2pt} \\ 
\scriptsize $^{10}$ Department Physics and Astronomy, University of Utah, 115 S 1400 E, Salt Lake City, UT 84112, USA\vspace*{-2pt} \\ 
\scriptsize $^{11}$ IRFU,CEA, Universit\'e Paris-Saclay, F-91191 Gif-sur-Yvette, France\vspace*{-2pt} \\ 
\scriptsize $^{12}$ Kavli Institute for Particle Astrophysics and Cosmology, Stanford University, 452 Lomita Mall, Stanford, CA 94305, USA\vspace*{-2pt} \\ 
\scriptsize $^{13}$ Aix Marseille Univ, CNRS, CNES, LAM, Marseille, France\vspace*{-2pt} \\ 
\scriptsize $^{14}$ Instituto de Ciencias F\'isicas, Universidad Nacional Aut\'onoma de M\'exico, Av. Universidad s/n, 62210 Cuernavaca, Mor., M\'exico\vspace*{-2pt} \\ 
\scriptsize $^{15}$ Institut  d'Estudis  Espacials  de  Catalunya  (IEEC),  E08034  Barcelona,  Spain\vspace*{-2pt} \\ 
\scriptsize $^{16}$ Max-Planck-Institut f\"ur Extraterrestrische Physik, Postfach 1312, Giessenbachstr., 85748 Garching bei M\"unchen, Germany\vspace*{-2pt} \\ 
\scriptsize $^{17}$ University of Oxford, Denys Wilkinson Building, Keble Road, Oxford, OX1 3RH, UK\vspace*{-2pt} \\ 
\scriptsize $^{18}$ Department of Physics and Astronomy, Sejong University, Seoul 143-747, Korea\vspace*{-2pt} \\ 
\scriptsize $^{19}$ Department of Astronomy and Astrophysics, The Pennsylvania State University, University Park, PA 16802, USA\vspace*{-2pt} \\ 
\scriptsize $^{20}$ Institute for Gravitation and the Cosmos, The Pennsylvania State University, University Park, PA 16802, USA\vspace*{-2pt} \\ 
\scriptsize $^{21}$ Center for Cosmology and Particle Physics, Department of Physics, New York University, New York, NY 10003, USA\vspace*{-2pt} \\ 
\scriptsize $^{22}$ School of Physics and Astronomy, University of St Andrews, St Andrews, KY16 9SS, UK\vspace*{-2pt} \\ 
\scriptsize $^{23}$ Instituto de F\'isica, Universidad Nacional Aut\'onoma de M\'exico, Apdo. Postal 20-364, Ciudad de M\'exico, M\'exico\vspace*{-2pt} \\ 
\scriptsize $^{24}$ National Astronomy Observatories, Chinese Academy of Science, Beijing, 100101, P.R. China\vspace*{-2pt} \\ 
\scriptsize $^{25}$ School of Astronomy and Space Science, University of Chinese Academy of Sciences, Beijing 100049, P.R.China\vspace*{-2pt} \\ 
}

\date{Last updated 2015 May 22; in original form 2013 September 5}

\pubyear{2015}

\begin{document}
\label{firstpage}
\pagerange{\pageref{firstpage}--\pageref{lastpage}}
\maketitle

\begin{abstract}
The completed extended Baryon Oscillation Spectroscopic Survey (eBOSS) catalogues contain redshifts of 344,080 quasars between $0.8<z<2.2$ covering $4808\deg^2$, 174,816 luminous-red galaxies over $0.6<z<1.0$ covering $4242\deg^2$ and 173,736 emission-line galaxies between $0.6<z<1.1$ covering $1170\deg^2$ in order to constrain the expansion history of the Universe and the growth rate of structure through clustering measurements. Mechanical limitations of the fibre-fed spectrograph on the Sloan telescope prevent two fibres being placed closer than $62\arcsec$, the fibre-collision scale, in a single pass of the instrument on the sky. These `fibre collisions' strongly correlate with the intrinsic clustering of targets and can bias measurements of the two-point correlation function resulting in a systematic error on the inferred values of the cosmological parameters. We combine the new techniques of pairwise-inverse-probability weighting (PIP) and the angular up-weighting (PIP+ANG) to correct the clustering measurements for the effect of fibre collisions. Using mock catalogues we show that our corrections provide unbiased measurements, within data precision, of both the projected correlation function $\rm{w_p}\left(r_p\right)$ and the multipoles $\xi^{(\ell=0,2,4)}\left(s\right)$ of the redshift-space correlation functions down to $0.1\mhmpc$, regardless of the tracer type. We apply the corrections to the eBOSS DR16 catalogues. We find that, on scales greater than $s\sim20\mhmpc$ for $\xi^\ell$, as used to make Baryon Acoustic Oscillation and large-scale Redshift-Space Distortion measurements, approximate methods such as Nearest-Neighbour up-weighting are sufficiently accurate given the statistical errors of the data. Using the PIP method, for the first time for a spectroscopic program of the Sloan Digital Sky Survey (SDSS) we are able to successfully access the 1-halo term in the 3D clustering measurements down to $\sim0.1\mhmpc$ scales. Our results will therefore allow studies that use the small-scale clustering measurements to strengthen the constraints on both cosmological parameters and the halo-occupation (HOD) distribution models.

\end{abstract}




\newpage

\section{Introduction}
The 3D distribution of galaxies in the Universe contains a wealth of information about its composition and dynamical evolution. One of the most efficient ways to access this information is provided by the spectroscopic galaxy redshift surveys. Spectroscopic redshift surveys over the past two decades \citep[]{york2000,drinkwater10,eisenstein11,guzzo14,blanton17} have thus played a crucial role in both constraining the standard cosmological model \citep[e.g.][]{percival01,cole05,eisenstein05,blake12,alam17} and probing the galaxy formation and evolution at different cosmic epochs \citep[e.g.][]{lewis02,chen13,krywult17}.

The advent of multi-object spectrographs (MOS) has allowed an unprecedented increase in the surveyed volume and catalogue size of spectroscopic redshift surveys allowing us to measure the cosmological observables with increasing precision. This requires an equivalent effort to correct for systematic issues that can potentially bias the measurements of the observables degrading the accuracy on estimates of physical parameters \citep{ross12,delatorre13}. In spectroscopic redshift surveys, based on multi-slit or multi-fibre spectrographs, one of these effects is caused by missing observations that result from the limitations of the hardware setup. In particular, the finite size of fibres (or slits) prevents any pair of targets closer than the fibre diameter to be observed simultaneously, usually referred to as the \emph{fibre collision} or \emph{close pair} problem \citep{ying98}. The effect is strongly correlated with the intrinsic clustering of targets as collisions occur in regions of high target density. The resulting systematic offset in the clustering measurements such as the two-point correlation function (2PCF) can mimic the effect of physical parameters such as galaxy bias $b$ or growth rate of structure $f\sigma_8$ introducing a systematic bias in the inferred values of these parameters \citep{pezzotta17}. Different techniques have been proposed in the past \citep[see e.g.][]{guo12,reid14,hahn17,zarrouk18,yang19,sunayama2020} to mitigate for missing observations in clustering measurements. Spectroscopic programs of the Sloan Digital Sky Survey have often previously used variations of the ``nearest neighbour'' (NN) technique to correct for the fibre collision problem. Here, either the redshift and classification of the nearest observed target is assigned to an un-observed one \citep{zehavi02,zehavi05}, or the nearest observed target is given the additional weight of the un-observed target. The Baryon Oscillation Spectroscopic Survey (BOSS) part of SDSS-III analyses \citep{reid16} adopted the latter variant, where the weight of an un-observed target was transferred to the closest observed target. The VImos Public Extragalactic Redshift Survey (VIPERS) employed the ``Target Sampling Rate'' (TSR) to quantify the selection probabilities. TSR is defined as the ratio between the number of observed and input targets in a rectangular region around an observed target where the size of the region was calibrated using realistic survey mock samples \citep{delatorre13}. The common factor of these techniques is that they do not correctly allow for the correlation between the observed and un-observed targets. For example, the up-weighting variant of the NN method ignores pairs between observed and un-observed targets, while the redshift assignment variant assumes that these pairs are all transverse to the line-of-sight.

In order to correctly allow for the correlation between the observed and un-observed targets, \cite{bianchi17} proposed a `pairwise-inverse probability' (PIP) weighting scheme. The technique calculates selection probabilities for each observed pair by generating multiple random survey realisations statistically equivalent to the actual observations, and then up-weighting each observed pair of targets by the reciprocal of this probability. The method is statistically unbiased under the assumption that no pair of targets has zero probability of being observed. \cite{percival17} extended the PIP technique under the assumption that the up-weighted pairs of a particular separation are statistically equivalent to those in the full sample. In this case, the PIP up-weighted angular pair counts can simply be scaled to match those in the full target sample. For Sloan telescope observations, this allows the full clustering signal to be recovered on all pair separations even if some regions are only covered by one observation. These techniques have been successfully tested for the multi-fibre spectroscopy adopted by the Dark Energy Spectroscopic Instrument (DESI) using mock samples \citep{bianchi18,smith19}. \cite{mohammad18} applied the PIP corrections to the final VIPERS dataset based on observations using the multi-slit VIsible Multi-Object Spectrograph (VIMOS) \citep{lefevre03}.

In this paper we apply the correction schemes presented in \cite{bianchi17} and \cite{percival17} to correct for missing observations affecting clustering measurements from the extended Baryon Oscillation Spectroscopic Survey (eBOSS) data \citep{dawson16}, one of the components of the SDSS-IV project \citep{blanton17}. eBOSS targeted three different samples: Emission-Line Galaxies (ELGs) between $0.6<z<1.1$ over $1170\deg^2$, Luminous-Red Galaxies (LRGs) between $0.6<z<1.0$ over $4242\deg^2$ and quasars (QSOs) split into `clustering' quasars between $0.8<z<2.2$ over $4808\deg^2$ and high redshift quasars for Lyman-$\alpha$ forest analyses. A number of studies have used clustering measurements from the early eBOSS data at relatively small scales \citep[see e.g.][]{laurent17,alam19,guo19}. However, none of these works have extended the analysis below $\sim 1 \mhmpc$ where fiber-collisions strongly affect the measured correlation function.

This study is part of a coordinated release of papers based on the final eBOSS catalogues. These include final eBOSS measurements of the baryon acoustic oscillations and redshift-space distortions in the clustering of luminous red galaxies \citep[LRG $0.6<z<1.0$;][]{bautista20,hector20}, emission line galaxies \citep[ELG $0.6<z<1.1$;][]{anand20,tamone20,arnaud20}, and quasars \citep[QSO $0.8<z<2.2$;][]{jiamin20,richard20}. Also part of this joint release are a set of papers describing the data catalogues \citep{ross20,lyke20}, mock catalogues \citep{lin20,zhao20} and N-body simulations for assessing systematic errors \citep{alam20,avila20,rossi20,smith20}. At the highest redshifts ($z>2.1$), the coordinated release of final eBOSS measurements includes measurements of BAO in the Lyman-$\alpha$ forest \citep{2020duMasdesBourbouxH}. \citet{eBOSS_Cosmology} presents the cosmological interpretation of these results combined with the final BOSS results and other probes\footnote{A description of eBOSS and links to all associated publications can be found here: \url{https://www.sdss.org/surveys/eboss/}}.

In Sec. \ref{sec:survey} we describe the main features of eBOSS catalogues, spectroscopic observations, details of the fibre assignment algorithm and the veto masks. In Sec. \ref{sec:randoms} we describe the random catalogues used to perform clustering measurements. We present the PIP and angular up-weighting schemes in Sec. \ref{sec:method}. Tests on mock catalogues are described in Sec. \ref{sec:tests} along with details on the survey mocks. We present the measurements from the final eBOSS catalogues in Sec. \ref{sec:results}. Finally, we summarise our results and draw conclusions in Sec. \ref{sec:conclusions}. 

Throughout this paper we adopt the same $\Lambda$CDM fiducial cosmology adopted in other eBOSS DR16 papers with parameters: $\Omega_m=0.31$, $\Omega_\Lambda=0.69$, $\Omega_bh^2=0.022$, $h=0.676$, $\sigma_8=0.8$ and $n_s=0.97$.

\section{Survey}\label{sec:survey}
 \begin{figure*}
    	\centering
		\includegraphics[scale=0.20]{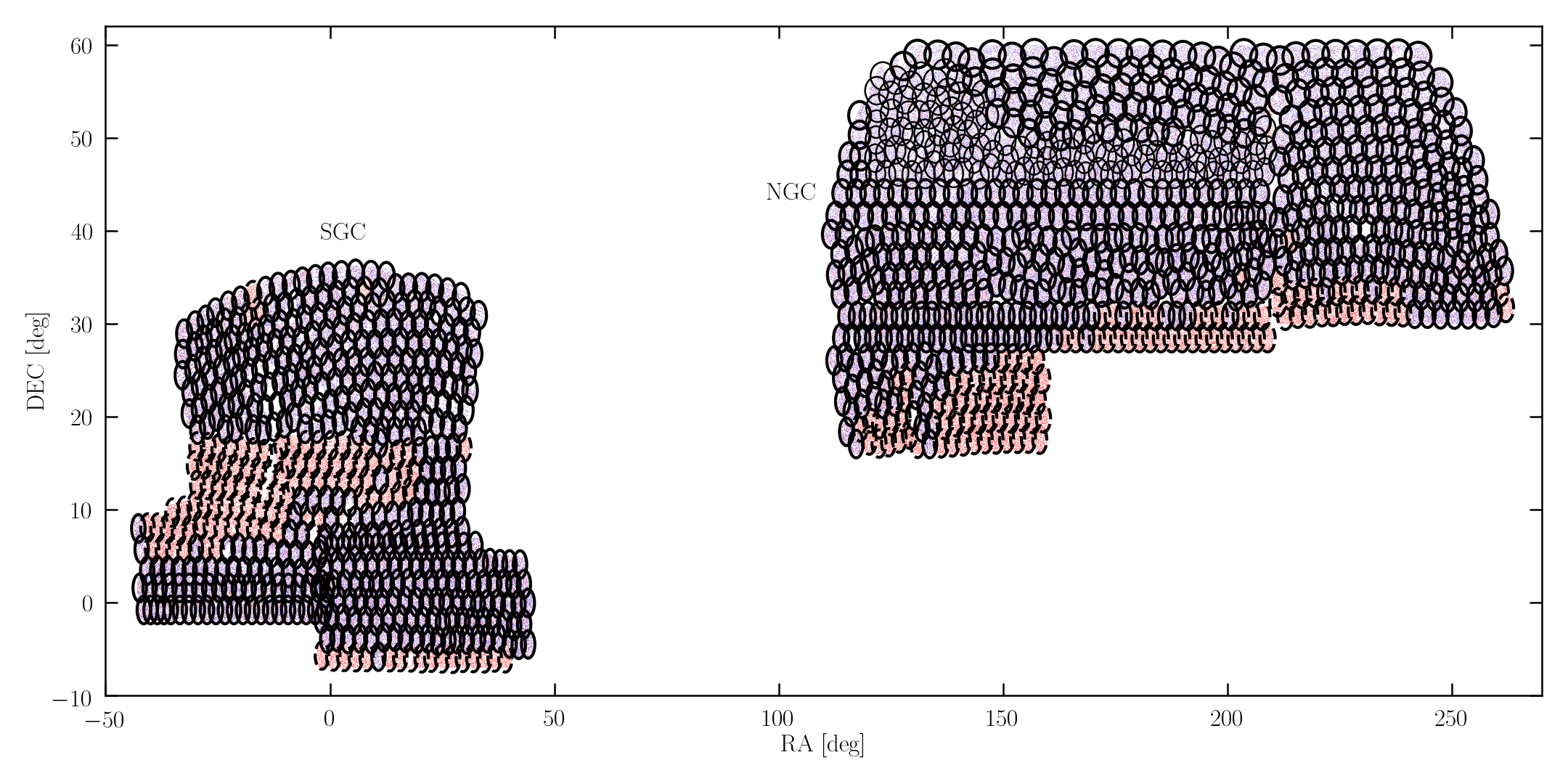}
		\caption{Sky coverage of the eBOSS DR16 QSO catalogue. Blue dots are targets included in the LSS catalogue used for the cosmological analyses. Red dots represent targets not included in the LSS catalogues due to a number of issues such as redshift failure, fibre collisions, veto masks, low survey completeness. Black thick (thin) ellipses show the positions of eBOSS (SEQUELS) plates tiled for subsequent observation. Solid ellipses show tiles that were observed, and dashed ellipses those that were not observed. eBOSS LRGs have the same large-scale window function as the quasars and hence are not shown.}\label{fig:tiling_qso}
	\end{figure*}
\begin{figure*}
    	\centering
		\includegraphics[scale=0.20]{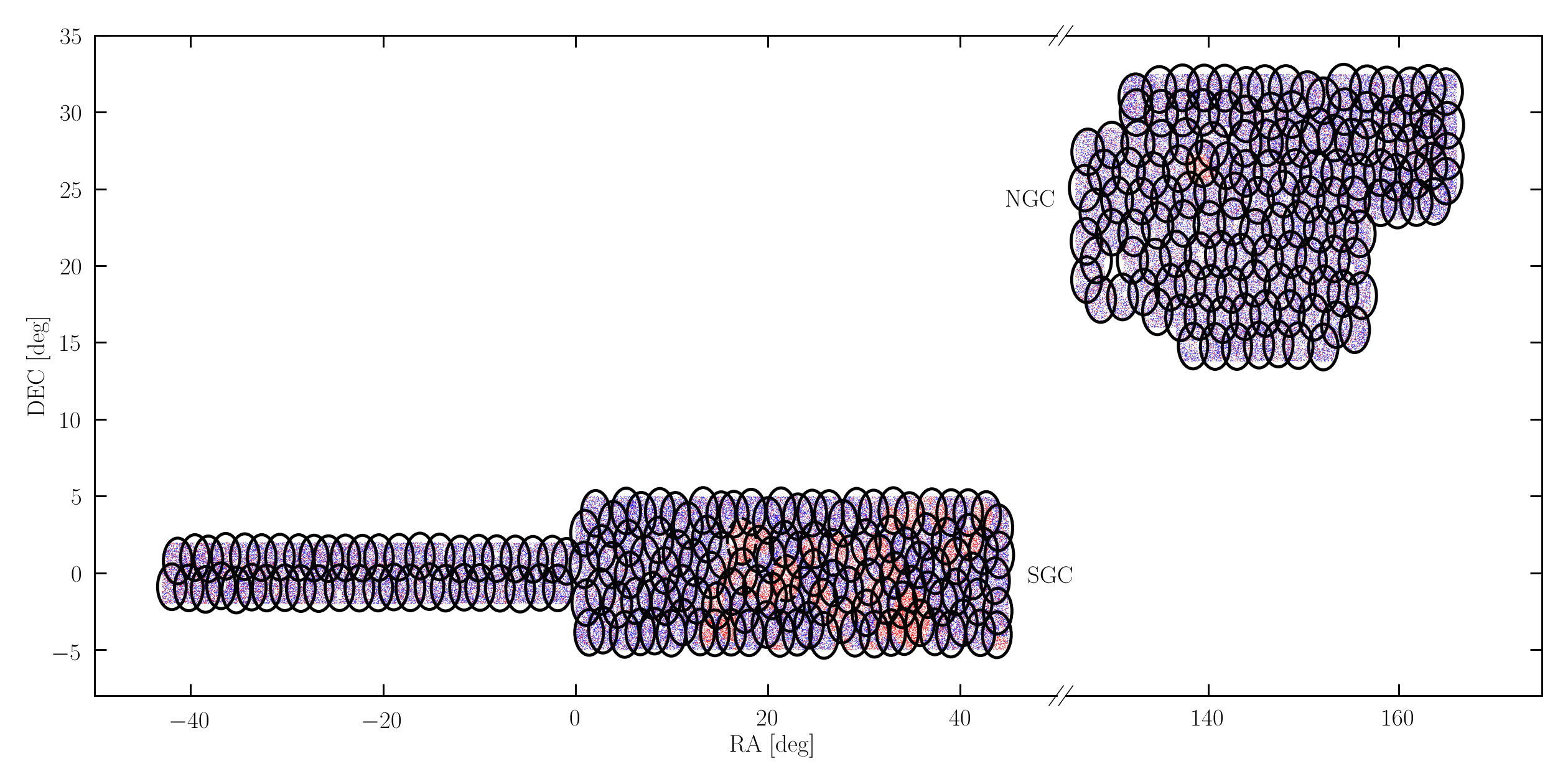}
		\caption{Same as in Fig. \ref{fig:tiling_qso} but here for the sample of ELGs.} \label{fig:tiling_elg}
	\end{figure*}

The extended Baryon Oscillation Spectroscopic Survey was conceived to build upon the remarkable achievements of BOSS, and design a complementary survey to higher redshift. As well as extending the sample of high redshift quasars for Lyman-$\alpha$ forest analyses and the well-known sample of luminous red galaxies from BOSS, it was designed to explore two new tracers: emission-line galaxies and quasars used as direct tracers of the large-scale structure (LSS). The survey design and target selection was driven by the goal of achieving a $\sim 1\%$ level precision on the measurements of the baryon acoustic oscillations (BAO). The target selection of different tracers is beyond the purpose of this paper and for the details of target selection we refer the reader to \cite{anand17} for the ELG sample, \cite{prakash16} for LRG catalogue and \cite{myers15} for QSOs. Detailed description of the survey design is provided in \cite{dawson16}, while LSS catalogue creation is outlined in \cite{ross20} for LRG and QSO targets and in \cite{anand20} for the ELG sample.

eBOSS collected spectra using the double-armed fibre-fed BOSS spectrograph \citep{smee13} at the 2.5m Sloan Telescope \citep{gunn06}. For each exposure, light is collected using optical fibres positioned in a pre-drilled aluminium plate placed at the focal plane of the telescope and transferred to the spectrographs. Each plate (or tile) covers $\sim7\deg^2$ and can accommodate 1000 fibres. Among these, 100 fibres are reserved for calibration targets with the remaining 900 dedicated for science targets. Each fibre and associated ferrule has a diameter of $62\arcsec$, the `fibre-collision' angle $\theta^{\rm{fc}}$. This puts a mechanical limitation on the minimum distances between two fibres on a single plate and prevents any pair of targets at smaller separation to be targeted simultaneously. However, multiple plates are allowed to partially overlap and the collision rate is therefore reduced in areas that are covered by more than one plate.

Luminous red galaxies and quasars shared the same plates as they were targeted simultaneously over $\sim5000\deg^2$ (Fig. \ref{fig:tiling_qso}) along with targets from the eBOSS sub-programs, namely the Spectroscopic IDentification of ERosita Sources (SPIDERS) and the Time-Domain Spectroscopic Survey (TDSS). A number of targets in the Legacy Quasar target class, identified as having secure redshifts already measured from BOSS and previous surveys, were not re-observed. Emission-line galaxies were observed using 300 dedicated plates over a limited area of $\sim1200\deg^2$ (Fig. \ref{fig:tiling_elg}) with a small fraction of fibres assigned to TDSS targets. The survey footprint is split into two regions denoted as the north- (NGC) and the south galactic cap (SGC) (see Fig. \ref{fig:tiling_qso} and Fig. \ref{fig:tiling_elg}). As part of SDSS, eBOSS uses bitmasks to store information about targets \footnote{\url{https://www.sdss.org/dr16/algorithms/bitmasks/}}. In eBOSS the \verb EBOSS_TARGET1 \ and \verb EBOSS_TARGET2 \ bitmasks are used to store the target types. The \verb EBOSS_TARGET1 \ bits that identify eBOSS `clustering' targets are \verb LRG1_WISE \ for luminous red galaxies and \verb QSO1_EBOSS_CORE \ for clustering quasars. \verb EBOSS_TARGET2 \ bits \verb ELG1_NGC \ and \verb ELG1_SGC \ are used to identify emission-line galaxies.

In this paper we use PIP weights for galaxies within the eBOSS DR16 LSS catalogues \citep{ross20,anand20}. These catalogues include only targets with good redshifts in the redshift range of interest and are restricted over the footprint where the survey completeness is greater than $50\%$. The survey completeness, denoted \verb COMP_BOSS , is defined as the ratio between the number of observed targets plus targets missed due to fibre-collisions and the total number of target candidates within a `sector'. A sector is defined as the region covered by a unique set of plates. The redshift range of the LSS catalogues is limited to $0.6< z< 1.0$ for luminous red galaxies, $0.8< z<2.2$ for quasars, and to $0.6< z<1.1$ for emission-line galaxies.

eBOSS DR16 catalogues of luminous red galaxies and quasars also include eBOSS targets observed as part of the Sloan extended quasar, ELG and LRG Survey (SEQUELS), which linked the BOSS and eBOSS experiments. These targets are gathered in two adjacent regions (or chunks, see Sec. \ref{sec:tiling_code}) denoted \verb boss214 \ and \verb boss217 \ in NGC. These regions were tiled using the software version and the priority system used for BOSS observations. 

\subsection{Fibre Assignment}\label{sec:tiling_code}

In eBOSS the selection of targets for spectroscopic observation was performed through the survey `tiling' process with the goal of maximising the number of targets that receive a fibre with a given number of tiles \citep{blanton03}. 

The fraction of targets of a given type that receive a fibre defines the tiling completeness. Collision groups are defined as a set of targets where each member of the group is in a fibre collision (i.e. with separation below $62\arcsec$) with at least one other member, such that they cannot all be observed within a single exposure. The tiling is performed to maximize the tiling efficiency (i.e., the fraction of science fibres assigned to targets) while reaching the desired densities and decollided completeness\footnote{The decollided set of targets consists of targets not in collision groups combined with colliding targets that can be assigned a fibre on a single plate \citep[see][]{dawson16}.} of various targets.

As described in \cite{blanton03} in detail, the tiling was performed in three steps: 
\begin{enumerate}
    \item first, the full survey footprint was divided into multiple chunks. In eBOSS different chunks were observed independently even if they partially overlap, i.e. targets in the chunks overlap regions can potentially be targeted in more than one chunk (See \cite{ross20} for details on how these areas are treated in the LSS catalogue creation);
    \item given the angular distribution of targets, tile centers in each chunk were initially drawn from a uniform covering of the celestial sphere which were then perturbed with respect to these initial positions to maximise the tiling completeness;
    \item finally, given the tiling solution from (ii) fibres were assigned to targets, eventually solving collisions between targets.
\end{enumerate}
Steps i-iii were adopted for the sufficiently large chunks. For chunks with narrow or small geometry, step (ii) in the procedure above was not able to produce an optimal configuration of tiles to reach the desired fibre efficiency and target density. For these chunks we manually placed evenly-located tiles for step (ii).

Given that different tracers were observed simultaneously, eBOSS observations adopted a tiered-priority system for assigning fibres to targets. A collision between two targets with different priorities is referred to as \textit{`knockout'} as opposed to fibre collisions that occur between targets of the same type. 

For the chunks with luminous red galaxies and quasar targets, fibres were assigned in two rounds: a) all non-LRG targets get maximum priority and receive fibres first requiring a $100\%$ tiling completeness for their decollided set; b) in the second round, remaining fibres are assigned to LRG targets at a lower priority. The requirement of $100\%$ decollided completeness for LRG targets was lifted since they were selected with a higher number density than the available fibres and they collide with targets with higher priority.\footnote{In fact, the tile centers of these LRG/QSO chunks were chosen using a downsampled set of the main LRG samples (by 15\%) in order to decrease the density of tiles; once the tile centers were decided, the fibre assignment pipeline was applied again to the entire sample.}  In the first round, collisions between non-LRG targets were resolved with following decreasing priorities: SPIDERS, TDSS, known quasars selected for re-observations, clustering quasars, variability-selected quasars, quasars from the Faint Images of the Radio Sky at Twenty-Centimeters (FIRST) survey and white dwarfs. The tiling density is on average one tile per 5 square degrees for chunks observed as part of eBOSS while it is approximately one tile per 4 square degrees for the two chunks observed as part of SEQUELS.

For the chunks dedicated to ELGs, the target densities were dominated by ELG targets in the first round with a small number of TDSS Few-Epoch Spectroscopy (FES) objects targeted at a density of $\sim 1\deg^{-2}$ at the same priority as ELG targets\footnote{Addititional TDSS targets were targeted on the same plates at lower priorities. The total density of TDSS targets that share plates with emission-line galaxies is $\sim 15 \deg^{-2}$.}. The tiling density of these chunks is on average one tile per around 4 square degrees.

\subsection{Veto Masks}

The portion of the survey area where spectroscopic observations are impossible is accounted for by different veto masks. Under the assumption that veto masks do not correlate with the target sample, their effect is to change the survey window function. As such these regions are masked from both data and random catalogues used for clustering measurements \citep{ross20,anand20}.

In particular, for LRG and QSO samples, areas contaminated by bright stars are masked using the `bright-star' veto mask. Other bright objects such as bright local galaxies and bright stars missed by the bright-star mask were also visually identified and masked using the `bright-object' veto mask \citep{reid16}. Regions with bad or no photometric observations are removed using the `bad-field' veto mask. No fibre can be placed within $92\arcsec$ of the center of each plate where a hole is drilled for the centerpost. These regions are part of the `centerpost' veto mask. A mask around infrared bright stars was also applied to the LRG target file before tiling. Knockouts are also assumed to be un-correlated with the target sample and are accounted for by applying the `collision-priority' veto mask. In regions covered by a single tile, the collision-priority mask removes any QSO within $62\arcsec$ of a TDSS or SPIDERS target. The LRG collision-priority mask removes any LRG within $62\arcsec$ from a non-LRG target, regardless of whether it lies in single-pass or multi-pass regions. This is a conservative approach motivated by the fact that, being targeted at the lowest priority, not all knocked-out LRGs are targeted in areas covered by more than one tile. For further details on the construction and properties of different veto masks we refer the reader to \cite{ross20}.

The sample of emission-line galaxies presents similar effects. However, veto masks for ELG sample were more complicated and built in form of pixelized masks. They account for issues related to different bright stars, systematics in the target selection and defects in the photometry. As anticipated in Sec. \ref{sec:tiling_code}, ELG targets share plates with some TDSS targets. In particular, TDSS FES-type targets have the same targeting priority as ELGs. To account for this a veto mask, similar to the collision-priority mask for LRG and QSO targets, is created around each TDSS-FES target. We refer the reader to \cite{anand20} for a detailed description of the veto masks for emission-line galaxies.

\subsection{Weights}

In redshift surveys the target selection is affected by different systematic effects that can alter the observed target density with respect to the true one \citep{ross12, delatorre13, bautista18, ross20}. Target selection from photometric data, fibre assignment and inaccurate redshift estimates are typical systematic effects that affect the observed target density in spectroscopic redshift surveys. Each systematic effect is corrected by applying a weight to each target. In eBOSS, targets in the LSS catalogues are assigned the following weights to correct for systematics or optimise the clustering measurements:
\begin{itemize}
    \item $w_{\rm{sys}}$ corrects for spurious fluctuations in the photometric target selection;
    \item $w_{\rm{cp}}$ is the standard correction to fibre collisions adopted in eBOSS cosmological analyses. In the following part of this paper we will refer to the $w_{\rm{cp}}$ weighting as the `CP' correction (or weighting) method;
    \item $w_{\rm{noz}}$ accounts for redshift failures;
    \item $w_{\rm{FKP}}$ are the standard FKP weights \citep{feldman94} used to minimise the variance of the measurement.
\end{itemize}
The CP correction is a variant of the standard NN method (see \cite{ross20}). In particular, $w_{\rm{CP}}$ weights are computed for collision groups where the weight of each target missed due to a fibre collision is equally distributed among the observed members of the group rather than assigning it only to its nearest neighbour as in the standard NN method. In this work we use the PIP weights as an alternative to $w_{\rm{cp}}$ keeping all other weights as in eq. \eqref{eq:weights}.

The overall standard weight is then:
    \begin{equation}
        w_{\rm{tot}} = w_{\rm{sys}}\times w_{\rm{noz}}\times w_{\rm{FKP}}\times w_{\rm{cp}}. \label{eq:weights}
    \end{equation}{}
The same weights listed in this section are also assigned to the objects in the random catalogues used to perform clustering measurements. However, the task of these weights for random points is to match the radial selection function of eBOSS targets rather than correcting for the related systematic effects. An exception is for $w_{\rm{sys}}$ weights for random points that are used to correct for survey completeness in each sector when the CP correction is adopted to correct for fibre collisions. We report in Sec. \ref{sec:randoms} the procedure used in \cite{ross20} and \cite{anand20} to assign weights to objects in the random catalogues.


\section{Random Catalogues}\label{sec:randoms}

For clustering measurements we need to compare the galaxy distribution to the expected distribution, or window function, in order to determine the overdensity. Given the complexities in the window function this is usually quantified using random catalogues, matching the angular and radial selection functions of data targets, coupled with a set of weights applied to the galaxies. Here we outline the main features of the random catalogues used in this paper. We refer the reader to the corresponding ELG \citep{anand20,arnaud20} and LRG/QSO \citep{ross20} catalogue papers for further details.

For all eBOSS tracers, random points are distributed homogeneously over the sky area covered by the survey and subsequently masked to match the footprint of the catalogues used for clustering analyses. This includes removing patterns inside vetoed regions and those excluded due to the low survey completeness (regions where \verb COMP_BOSS \ drops below $50\%$). 

The next step is to assign redshifts to random points making sure they accurately reproduce the radial selection function of eBOSS targets. This is done in different ways for different tracers:
\begin{itemize}
    \item[-] In the case of LRG and QSO samples, to ensure that random catalogues match the effective radial distribution of targets, their redshift and weights ($w_{\rm{sys}}$,$w_{\rm{cp}}$,$w_{\rm{noz}}$,$w_{\rm{FKP}}$) are drawn from a randomly selected galaxy within the sample.
    \item[-] The radial selection function of ELG targets depends on the imaging depth. To account for this effect, survey area is first divided in sub-regions of approximately equal \emph{grz} imaging depth. The redshift and $w_{\rm{FKP}}$ for a random point in a given sub-regions is then drawn from a randomly selected ELG target in the same sub-region with a probability proportional to $w_{\rm{sys}}\times w_{\rm{cp}}\times w_{\rm{noz}}$. A normalization factor is included in the $w_{\rm{sys}}$ weights of random points to assure that the ratio between weighted sums of random points and ELG targets is constant among different chunks while $w_{\rm{cp}}$ and $w_{\rm{noz}}$ are set to unity.
\end{itemize}

The standard CP correction for the fibre collision adopted in eBOSS cosmological analyses, a variation of the nearest-neighbour (NN) weighting (see Sec. \ref{sec:method}), does not account for the sector-to-sector variation in the survey completeness (\verb COMP_BOSS). \ This requires either up-weighting target samples by the survey completeness or down-weighting the randoms by the same quantity. The last option is adopted by embedding a \verb COMP_BOSS \ factor in the $w_{\rm{sys}}$ weights for random points \citep{ross20,anand20}. Consequently, a fraction of legacy quasars in each sector is removed to match the completeness of eBOSS quasars. We make use of these random catalogues when correcting for fibre collisions with the modified NN weighting scheme adopted in eBOSS.

The PIP weights, on the other hand, are inferred by re-running the eBOSS fibre assignment algorithm. As such they already account for the survey completeness since this lowers the probability of any target to get a fibre in sectors where the number of fibres is lower than the number of decollided targets. Consequently, no correction is required for the randoms, and in effect the PIP weights leave an isotropic expected galaxy distribution within the survey mask. We thus remove the completeness factor from the $w_{\rm{sys}}$ weights for randoms when using the PIP corrections for fibre collisions. Furthermore, this also allows us to bring back into the LSS sample all legacy quasars that were down-sampled to match the survey completeness.

\section{Methodology}\label{sec:method}
In this section we present the details of our analysis methodology. Before describing the PIP up-weighting method, we first recap how standard measurements of the two-point correlation function are performed as this sets the scene for the PIP up-weighting method. We then present the main features of the PIP and angular up-weighting schemes and the specific way that target selection probabilities are inferred for eBOSS galaxy samples.

\subsection{Measurement of the correlation function}
We adopt the widely used least-biased and least-variance Landy-Szalay estimator \citep{landy93} to measure the two-point correlation functions:
    \begin{equation}
        \xi(\mathbf{s}) = \frac{DD(\mathbf{s})-2DR(\mathbf{s})}{RR(\mathbf{s})}+1, \label{eq:LS}
    \end{equation}
where $DD$, $DR$ and $RR$ are the data-data, data-random and random-random pair counts normalised to the total number of corresponding weighted pairs and $\mathbf{s}$ is the pair separation vector in redshift space, i.e. when distances are inferred though the observed redshifts.

We measure two types of correlation functions:

\begin{itemize}
    \item[-]  the projected correlation function $\mwp$ that is commonly used in literature to constrain the Halo Occupation Distribution (HOD) models,
    \begin{equation}
        \mwp = 2\int_0^{\pi_{\rm{max}}}\xi\left(r_p,\pi\right)d\pi\ . \label{eq:wp}
    \end{equation}
\item[-] The multipoles $\xi^{(\ell)}$ of the redshift space 3D correlation function, widely used to measure and model anisotropic 3D clustering in redshift space,
    \begin{equation}
        \xi^{(\ell)}(s) = \left(2\ell+1\right)\int_{0}^{+1}\xi(s,\mu)L_\ell(\mu)d\mu. \label{eq:mps}
    \end{equation}{}
\end{itemize}

In eq. \eqref{eq:wp}, $r_p$ and $\pi$ are the transverse and parallel to the line-of-sight components of the pair separations $\mathbf{s}$. The integral in eq. \eqref{eq:wp} is truncated at $\pi_{\rm{max}}=80\mhmpc$ as measurements at larger scales are noise dominated. Given the angular coordinates $\left(\rm{RA},\rm{DEC}\right)$ and redshifts $z$ we compute $r_p$ and $\pi$ following \cite{fisher94}. In eq. \eqref{eq:mps}, the angle-averaged pair separation is $s^2=r_p^2+\pi^2$ and the cosine of the angle between pair separation and line-of-sight is $\mu=\pi/s$.

In order to resolve the clustering at sub-$\mhmpc$ scales we use a logarithmic binning for the pair separation $s$ and its transverse to the line-of-sight component $r_p$
    \begin{equation}
        \log s_{i+1} = \log s_i+\Delta s_{\log},
    \end{equation}{\label{eq:log_binning}}
where $\Delta s_{\log}=0.1$. The logarithmic mean of the bin edges is used as the sampling point \citep{mohammad18}. The line-of-sight component $\pi$ of the pair separation is binned using a $1\mhmpc$ linear binning. When computing the multipoles of the two-point correlation function we divide $\mu$ between $\left[0,1\right]$ in 200 linear bins.

\subsection{Pairwise-Inverse Probability (PIP) Weighting}

The PIP weight for a given pair is defined as the inverse of the probability of it being targeted within an ensemble set of possible realisations of the survey, which includes all possible pairs of targets, and from which the actual realisation of the survey undertaken can be considered to be drawn at random. The selection probabilities depend strongly on the particular fibre assignment algorithm adopted to select targets from a parent photometric catalogue for the spectroscopic follow-up, and are therefore difficult to model except by rerunning the actual algorithm adopted. We thus rely on inferring the selection probabilities by generating multiple replicas of the survey target selection, changing the "random seed" for each run such that different choices are made in which target to select for follow-up spectroscopy (see Sec. \ref{sec:survey_realizations}). The inverse probability is then simply estimated as the number of realisations $N_{runs}$ in which a given pair could have been targeted divided by the number of times it was actually targeted \citep[see][for a discussion about inverse-probability estimators; specifically, following the nomenclature introduced in that work, we adopt the zero-truncated estimator]{bianchi19}. The PIP correction gives unbiased measurements of the two-point correlation function provided that there are no pairs with zero probability of being targeted in the ensemble of survey realisations. 

Following \cite{bianchi17}, rather than storing pairwise weights for individual pairs we store what are referred to as bitwise weights $w_i^{(b)}$ for each target. These are simply binary arrays of length $N_{runs}$ where each bit (either one or zero) represents the outcome of the corresponding fibre assignment run for target $i$ (either this target is, or is not included in run $b$). Bitwise weights are then combined \emph{``on the fly''} to compute the pairwise weights between target $m$ and target $n$ as:
\begin{equation}
    w_{mn} = \frac{N_{runs}}{\popcnt \left[w_m^{(b)}\&w_n^{(b)}\right]}.\label{eq:pip}
\end{equation}
In eq. \eqref{eq:pip} $\popcnt$ and $\&$ are standard bitwise operators. In particular, $\popcnt$ is the `population count' operator that given an array, in this case a bit sequence of 0 and 1, returns the number of elements different than 0. In eq. \eqref{eq:pip} $\&$ is the bitwise `\verb and ' that, given two arrays of equal length, performs the logical `AND' operation on each pair of the corresponding bits and returns the result as an array with length equal to that of input arrays. The weights $w_m$ for individual targets, called individual-inverse-probability (IIP) weights, can be calculated simply by replacing $m=n$ in eq. \eqref{eq:pip}. The same random catalogue is valid for all fibre assignment runs, and so the pair counts in eq. \eqref{eq:LS} are now:
\begin{equation}{\label{eq:pip_pairs}}
\left.\begin{aligned}
    DD(\vec{s}) &= \sum_{\vec{x}_m - \vec{x}_n \approx \vec{s}} \mathrm{w}_{mn}w^{'}_{\rm{tot,m}}w^{'}_{\rm{tot,n}} \ ,\\
    DR(\vec{s}) &= \sum_{\vec{x}_m - \vec{y}_n \approx \vec{s}} \mathrm{w}_{m}w^{'}_{\rm{tot,m}}w_{\rm{tot,n}} \ ,
    \end{aligned}\right.
\end{equation}
In eq. \eqref{eq:pip_pairs} $w^{'}_{\rm{tot}}=w_{\rm{sys}}\times w_{\rm{noz}}\times w_{\rm{FKP}}$ and $w_{mn}$ and $w_m$ are PIP and IIP weights, respectively. The $RR$ pairs are computed using the overall weights in eq. \eqref{eq:weights}.

\subsection{Angular Up-weighting}

The PIP weighting scheme is unbiased only if there are no pairs with zero selection probability. This is not the case for pairs with separations below the fibre-collision scale that fall in the single pass regions of the survey. Indeed these pairs are systematically missed regardless of the number of survey realisations used to infer the selection probabilities since at least one of the two targets cannot be observed. However, a fraction of colliding pairs are targeted in regions where two or more tiles overlap. Under the assumption that the set of un-observed pairs is statistically equivalent to the set of observed pairs, we can use the angular up-weighting scheme proposed in \cite{percival17} to recover the small-scale clustering. Although a reasonable assumption, survey designs may produce scenarios where this ansatz is not valid. In eBOSS, to maximise the targeting efficiency, areas where tiles overlap correlate to some extent with the regions of high targeting density. This can introduce a bias in the measurements from eBOSS DR16 catalogues shown in Fig. \ref{fig:wp_lrg_data}-\ref{fig:mps_elg_data} that cannot be corrected using methods available in literature.

The angular up-weighting is performed both on the $DD$ and $DR$ pair counts. Eq. \eqref{eq:pip_pairs} then becomes,
\begin{equation}{\label{eq:pip+ang}}
\left.\begin{aligned}
    DD(\vec{s}) &= \sum_{\substack{\vec{x}_m - \vec{x}_n \approx \vec{s}\\\vec{u}_m\cdot \vec{u}_n\approx\cos{\theta}}} \mathrm{w}_{mn}w^{'}_{\rm{tot,m}}w^{'}_{\rm{tot,n}}\times \wdd \ ,\\
    DR(\vec{s}) &= \sum_{\substack{\vec{x}_m - \vec{y}_n \approx \vec{s}\\\vec{u}_m\cdot \vec{v}_n\approx\cos{\theta}}} \mathrm{w}_{m}w^{'}_{\rm{tot,m}}w_{\rm{tot,n}}\times\wdr \ ,
    \end{aligned}\right.
\end{equation}
where $\vec{u}=\vec{x}/x$. The angular weights $\wdd$ and $\wdr$ in eq. \eqref{eq:ang_weights}, used to up-weight $DD$ and $DR$ pair counts respectively, are defined as,
\begin{equation}{\label{eq:ang_weights}}
\left.\begin{aligned}
    w_{\rm{ang}}^{\rm{DD}}(\theta) &= \frac{DD^{\rm{par}}\left(\theta\right)}{DD^{\rm{fib}}_{\rm{PIP}}\left(\theta\right)} \ ,\\
    w_{\rm{ang}}^{\rm{DR}}(\theta) &= \frac{DR^{\rm{par}}\left(\theta\right)}{DR^{\rm{fib}}_{\rm{IIP}}\left(\theta\right)} \ .
    \end{aligned}\right.
\end{equation}
The superscripts -$\rm{par}$ and -$\rm{fib}$ in eq. \eqref{eq:ang_weights} denote pairs of targets from the reference parent sample and pairs of targets that receive fibres, respectively. The subscript $\rm{PIP}$ and $\rm{IIP}$ denote the fact that the pair counts are up-weighted using the pairwise-inverse-probability weights (PIP) or their counterpart, the individual-inverse-probability weights (IIP), for the $DR$ pairs. In the following part of this paper we will use the abbreviation PIP+ANG to refer to the overall weighting outlined in eq. \eqref{eq:pip+ang}.

The angular weights derived in \cite{percival17} correct for the geometrical selection given by the survey targeting strategy. Because fiber assignment is independent of the properties upon which one normally selects sub-samples, eq. \eqref{eq:ang_weights} can therefore be applied to any sub-sample of the parent catalogue selected e.g. by colour or redshift. This is demonstrated by \cite{mohammad18}, where angular weights derived using the VIPERS parent sample were successfully applied to correct for fiber assignment for sub-samples selected in two different redshift bins. Angular up-weighting in the framework of the generalized inverse-probability weighting is discussed in \cite{bianchi19}.

 \begin{figure}
    	\centering
		\includegraphics[scale=0.09]{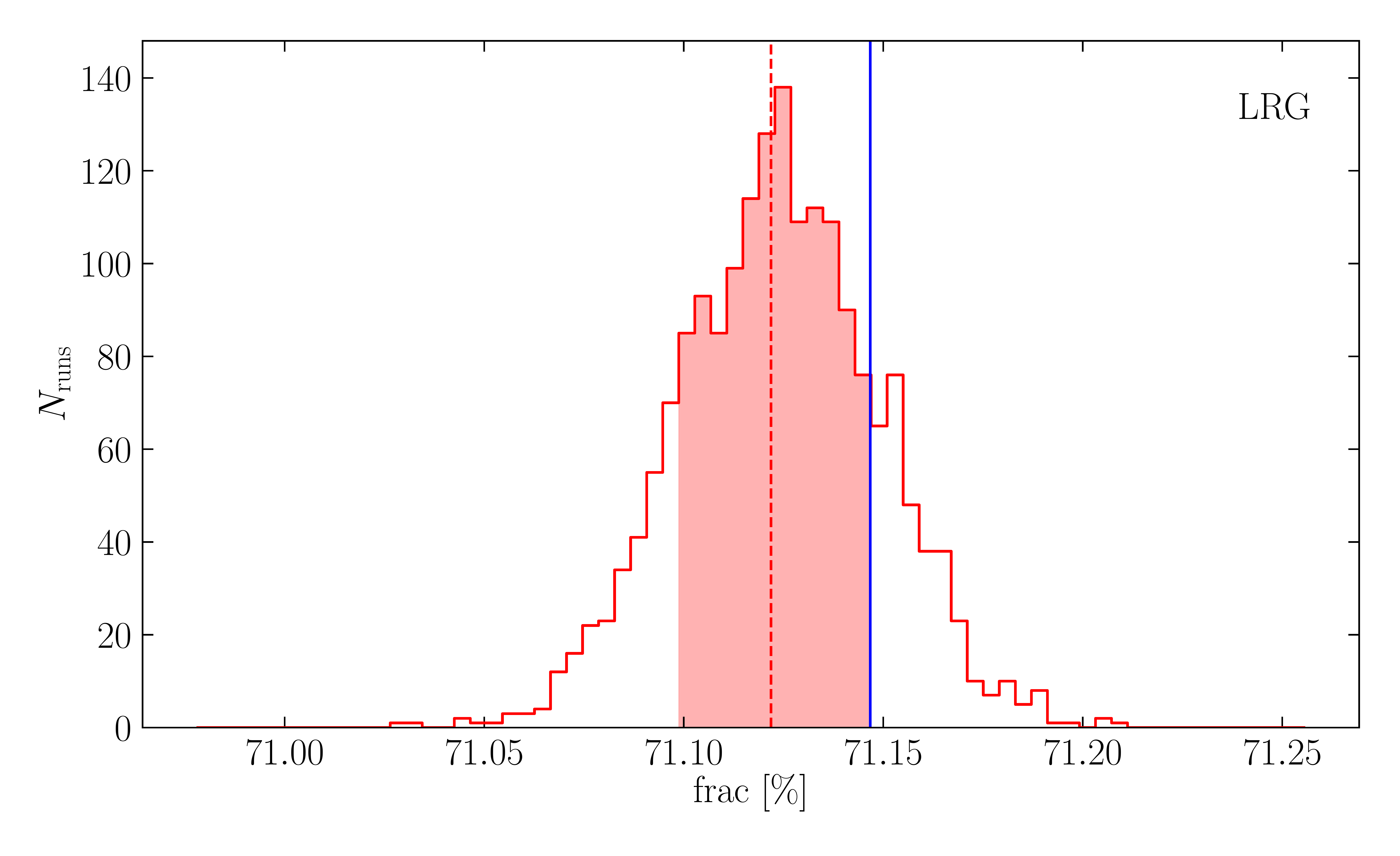}
		\caption{The distribution of the fraction of targets that get a fibre among 1860 fibre assignment runs on the LRG DR16 catalogue. The vertical red dashed line shows the mean of the distribution while the vertical shaded band represents the standard deviation. The vertical blue line shows the fraction of targets that received a fibre for the actual eBOSS observation.} \label{fig:tiling_stat_lrg}
	\end{figure}

 \begin{figure}
    	\centering
		\includegraphics[scale=0.09]{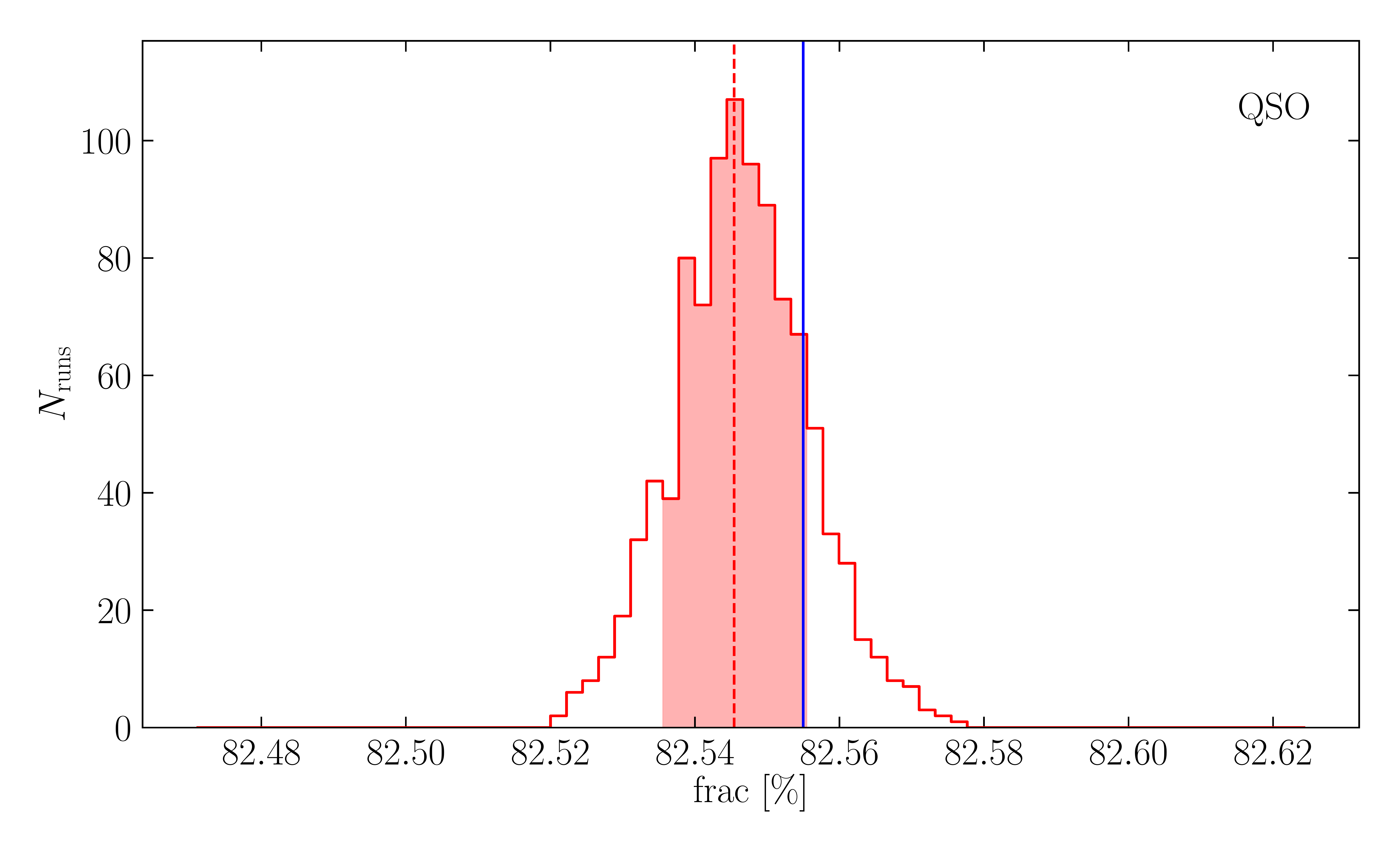}
		\caption{As in Fig. \ref{fig:tiling_stat_lrg}, here for the catalogue of quasars.} \label{fig:tiling_stat_qso}
	\end{figure}

 \begin{figure}
    	\centering
		\includegraphics[scale=0.09]{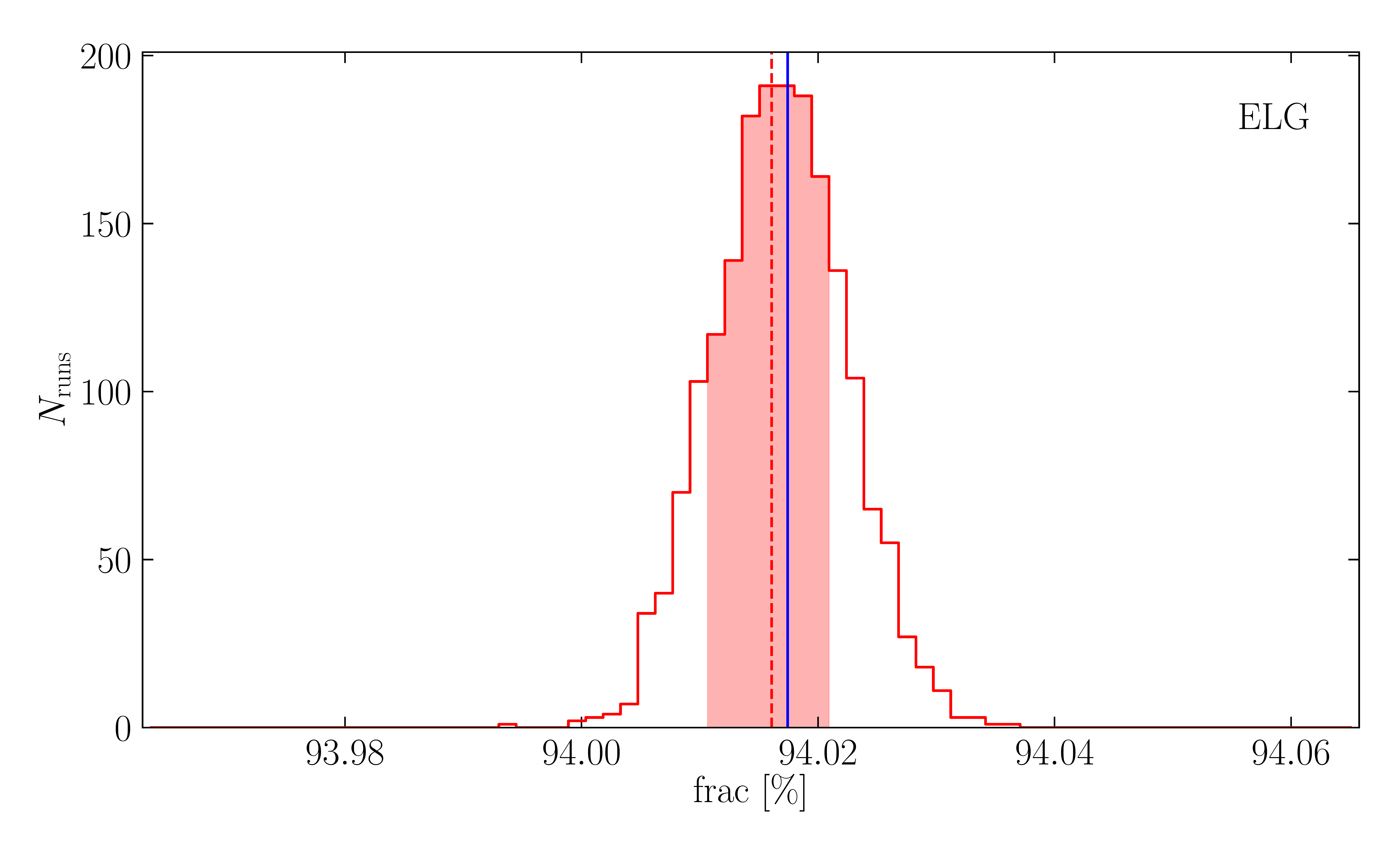}
		\caption{As in Fig. \ref{fig:tiling_stat_lrg}, here for the catalogue of emission-line galaxies.} \label{fig:tiling_stat_elg}
	\end{figure}

\subsection{Survey Realisations}\label{sec:survey_realizations}

To infer realistic selection probabilities, we generate multiple random realisations of the survey that are statistically equivalent to the actual observations. The eBOSS tiling algorithm resolves collisions between targets in a random fashion by means of a random number generator, except for targets in collision groups with more than two objects. For targets within these collision groups, the algorithm performs a procedure designed to optimise the number of fibres allocated to targets \citep{blanton03}. As we will see later, this causes some problems for the PIP weights close to the fibre collision scale, as a result of there being zero probability pairs. However, these effects are significantly below the noise level for any single realisation of the survey.

We generate multiple survey realisations by re-running the eBOSS fibre-assignment algorithm many times, changing the random seed used to initiate the random number generator each time. However, this feature is hard-coded in a large IDL-based software package that consists of several codes that optimize the fibre assignment and run sanity tests. Furthermore, the default setup only allows a single survey realisation to be generated at a time. For a processing time of $\sim1$-$2\mathrm{hr}$ this presents the main limitation for our purpose where thousands of survey realisations are needed for real datasets and for a relatively large number of mock samples. We have thus modified the relevant components of the eBOSS tiling software package in order to generate multiple survey realisations simultaneously by running it in parallel on a multi-processor computer cluster. In order to ease this task for future surveys we suggest implementing target selection algorithms in well-documented packages written in a fast and widely used programming language.

In re-running the fibre assignment, we keep the spatial distribution of tiles across all fibre-assignment runs fixed, matching the distribution used in eBOSS observations. The accuracy of the selection probabilities calculated in this way depends on the number of survey realisations used. We use a total of 1860 fibre assignment runs to infer the selection probabilities when using eBOSS DR16 catalogues. The first of these runs corresponds to that used for the actual eBOSS observations. The outcome of these runs is stored in bitwise weights that are then used to compute the PIP weights and correct the pair counts following eq. \eqref{eq:pip}-\eqref{eq:pip+ang}. Figures \ref{fig:tiling_stat_lrg}, \ref{fig:tiling_stat_qso} and \ref{fig:tiling_stat_elg} show the distribution of the fraction of targets that receive a fibre among these 1860 survey realisations (red histograms) and highlight the specific fibre assignment realisation that was used for eBOSS observations (vertical blue lines). The  run used for eBOSS spectroscopic observations (vertical blue line) shows a fraction of targets with fibre higher than the mean of the distribution, but within 1-$\sigma$ of the mean. For individual mocks the number of fibre assignment runs is limited to 310, a fair compromise between the accuracy of selection probabilities and the computation time. 

As shown in Fig. \ref{fig:tiling_qso}, a small fraction of plates were not observed even though they were included in the process of fibre assignment. Targets that were assigned a fibre from one of the un-observed plates thus contribute to lowering the survey completeness on the edges of the observed area. To correctly account for this, we run the fibre assignment using the full set of tiles and, once the fibres are assigned, we flag all targets that get a fibre from one of the un-observed plates as missed.

\section{Validation using mock catalogues}\label{sec:tests}
We use the Effective Zel'dovich mock samples \citep[EZmocks,][]{zhao20} to assess the performance of the PIP and angular up-weighting schemes. These mocks are each built from a Gaussian random field in a 5$\mhgpcc$ cubic box assuming an initial power spectrum and geometry matching a flat $\Lambda$CDM fiducial cosmology with parameters $\Omega_m=0.307115$, $\Omega_b=0.048206$, $h=0.6777$, $\sigma_8=0.8225$, $ns=0.9611$. The matter particles are displaced from their initial to final positions using the Zel'dovich approximation. The tracer number density is calculated using the matter density field and assuming a bias function that accounts for critical density required to form gravitationally bound structures, and for the stochasticity in the halo bias relation. The bias relation is calibrated using the real eBOSS dataset to match the clustering on linear and mildly non-linear scales. Redshift-space distortions are added by means of a linear term calculated using the Zel'dovich approximation that accounts for the bulk flows. Non-linear motions are included through an isotropic Gaussian motion added to the linear component. Mocks are then cut according to the survey geometry removing areas not included in the DR16 LSS catalogues. We refer the reader to \cite{zhao20} for a detailed description of the mock construction.

In this work we limit our analyses to only 100 mocks since running eBOSS fibre assignment on each mock catalogue requires a significant amount of computing time. The EZmocks made available to the eBOSS team are designed to match the LSS catalogues, in order to facilitate the cosmological analyses \citep{zhao20, arnaud20, jiamin20, richard20, hector20, anand20}. Among other effects, fibre collisions are emulated in these mocks in an approximate way, removing objects that do not receive a fibre, to reproduce the effect observed in eBOSS. Therefore this set of mocks is not suitable for the purpose of this work where we need to assess realistic selection probabilities that requires processing each mock catalogue through the same eBOSS tiling package used for actual observations. We thus use "raw" EZmocks, which have the same angular footprint as that of the eBOSS samples used for the cosmological analyses, and exhibit a flat radial selection function over the redshift range covered by the eBOSS DR16 LSS catalogues. We now describe the manipulations required in order to convert these mocks to match the eBOSS target samples.

\subsection{Adding Contaminants}\label{sec:contaminants}
Raw mocks were built to reproduce the clustering of the eBOSS DR16 LSS catalogues. However, they significantly differ from real data catalogues, used in the survey tiling, in the radial selection function and number density. We modify raw mocks to make them as realistic as possible and match the features of the data catalogues as following:
\begin{enumerate}
    \item all objects in the raw mocks are flagged as eBOSS clustering targets. In particular, LRGs and QSOs are assigned \verb EBOSS_TARGET1 \ bits \verb LRG1_WISE \ and \verb QSO1_EBOSS_CORE, respectively, while ELGs are flagged with \verb EBOSS_TARGET2 \ bits \verb ELG1_NGC \ (in NGC) and \verb ELG1_SGC \ (in SGC);
    \item mocks are then down-sampled to match the radial distribution of eBOSS clustering targets. In order to replicate the number density of targets we match the number of mock targets to the weighted number of eBOSS targets in narrow redshift bins;
    \item eBOSS data targets with un-known redshifts or redshifts outside the mocks redshift range are added to each mock;
    \item spectroscopically confirmed stars initially misidentified as eBOSS targets in data are added as contaminants to the mock catalogues;
    \item lack of mock targets in a posteriori vetoed regions, i.e. vetoed after survey tiling and spectroscopic observations, is compensated by adding targets from the eBOSS data in these regions to the mocks;
    \item a set of ancillary targets were observed simultaneously with eBOSS targets. When running the fibre assignment on mocks we supplement the mock catalogues resulting from steps i-v with the same ancillary target sample used in real observations. However, in data catalogues there is a strong correlation between ancillary targets and QSO catalogue. Namely $\sim 50\%$ of ancillary targets are within less than $2\arcsec$ of a quasar. These cases were treated as duplicates in eBOSS observations and they lower the number of fibres required to target quasar and ancillary samples. In mocks this fraction decreases to below $\sim 10\%$ since the correlation occurs mainly between ancillary targets and data targets that are added to mocks as contaminants in (iii)-(v). This has an effect of leaving a small number of fibres for mock LRG targets and therefore degrading the completeness of LRG sample targeted at a lower priority. To improve the completeness of mock LRG samples we randomly down-sample $\sim50\%$ of ancillary targets in chunks used for targeting LRG and QSO;
    \item finally, eBOSS `clustering' quasars that have known redshifts are not re-observed in eBOSS. These constitute $\sim 21\%\ (13\%)$ of eBOSS \verb QSO1_EBOSS_CORE  sample in NGC (SGC) over the redshift range covered by mocks. We thus flag the same fractions of mock quasars as known. As such these objects are not candidates to receive a fibre.
\end{enumerate}\label{en:mock_mod}

\begin{figure}
    	\centering
		\includegraphics[scale=0.09]{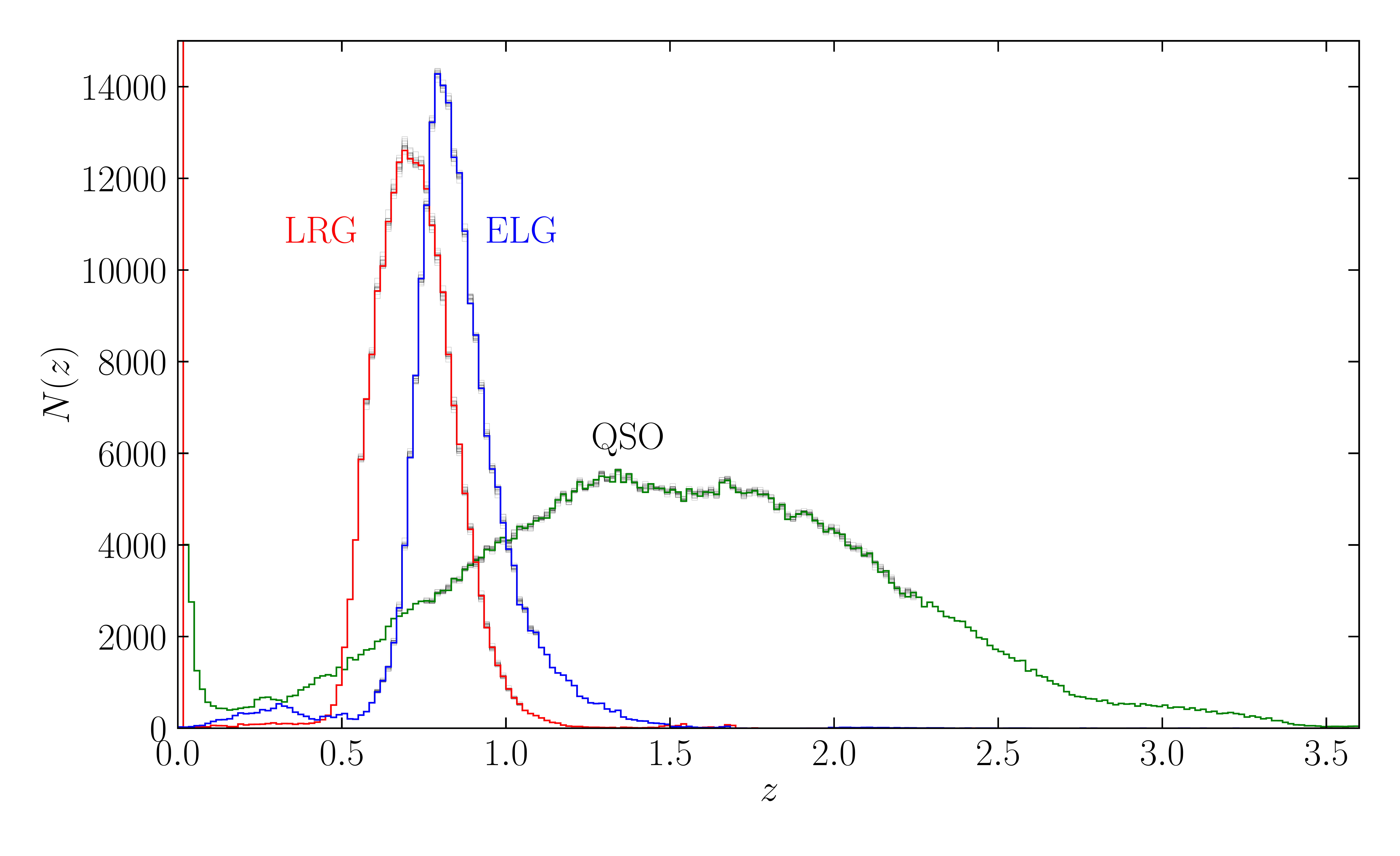}
		\caption{Weighted number of eBOSS DR16 LRG (red histogram), QSO (green histogram) and ELG (blue histogram) targets as a function of redshift $z$. The same quantities from 100 mocks are shown by the grey lines although they are not visually distinguishable from the coloured lines due to the tiny difference between data and mocks $N(z)$. Mock and data $N(z)$ histograms perfectly overlap at redshifts not covered by mock catalogues as all targets at these redshifts are taken from data catalogues (see Sec. \ref{sec:tests}).}\label{fig:Nz_lrg-qso}
\end{figure}

The redshift distribution of the modified mocks (not accounting for contaminants added in steps iii-v) is shown in Fig. \ref{fig:Nz_lrg-qso} along with the same quantity from eBOSS DR16 catalogues. The total number of targets in mocks differs, on average, from those in data catalogues by less than $\sim 5\%$.

We implement steps i-vii in order to make the fibre collisions in the mocks as close as possible to the real data. However, we are limited by the fact that EZmock catalogues are built using approximate prescriptions and are not based on the outcome of a full N-body simulation. As such it is inevitable that mock and data will present different small-scale clustering and consequently will have different collision rates. Nevertheless, this does not affect our conclusions as our tests are performed in order to demonstrate that we can recover the correct small-scale clustering without bias from the fibre collision issue. We therefore need to compare the clustering recovered from the mocks to the true clustering of the mocks, rather than to that of the data. It is worth noting that the three tracers show different small-scale clustering from each other as well as from the data, providing breadth to the tests performed.

Finally we process mocks obtained from steps i-vii through the eBOSS tiling code to implement the realistic fibre assignment. In running the fibre assignment on mock samples we follow the same procedure described in Sec. \ref{sec:tiling_code}. In particular, the full survey area is split into multiple chunks, identical to those used for eBOSS data catalogues and we use the eBOSS tiered-priority system \citep{dawson16} to solve collisions between different types of targets. In the following we will refer to mocks resulting from step i-ii as ``parent'' mocks while we will use the term ``spectroscopic'' mocks to refer to the corresponding sub-samples of mock targets that were assigned a fibre.

We do not include systematics other than fibre collisions in our parent and spectroscopic mocks. This choice is motivated by the fact that issues such as systematics in the photometric target selection and redshift failures are correlated with different galaxy properties \citep{ross12,scodeggio18,ross20,anand20} and  extremely difficult to accurately reproduce in simulated datasets. We therefore set $w_{\rm{sys}}=w_{\rm{noz}}=1$. The FKP weights are computed as $w_{\rm{FKP}} = 1/(1+\bar{n}(z)P_0)$ where $\bar{n}(z)$ is the mean number density of mock targets and $P_0=4000\mhmpcc$ for ELG \citep{anand20}, $P_0=6000\mhmpcc$ for QSO and $P_0=10000\mhmpcc$ for LRG \citep{ross20}. We infer the PIP weights through 310 fibre assignment runs on each mock.

The difference in the small-scale clustering between mocks and data also affects the implementation of the angular up-weighting for the mock catalogues. The key requirement for the angular up-weighting to be unbiased is that the sample of observed targets (labelled -fib in eq. \eqref{eq:pip+ang}), used for the clustering measurements, is statistically equivalent to the parent catalogue (labelled -par in eq. \eqref{eq:pip+ang}). In the case of mocks the `spectroscopic' mock sample is significantly different, in terms of small-scale clustering, than the mock catalogue used for fibre assignment since the latter one contains contaminants from eBOSS target catalogues added in steps iii-v. We therefore use mocks obtained from steps i-ii as the reference parent samples (discarding any target added from the eBOSS target catalogue) to compute quantities labelled with -$\rm{par}$ and the corresponding spectroscopic mock samples as the targeted samples to compute quantities labelled with -$\rm{fib}$ in eq. \eqref{eq:pip+ang}.

The goal of this section is to perform robustness tests for the novel PIP and PIP+ANG up-weighting scheme. We present an indirect test of the validity of the standard CP correction for fibre collisions adopted in eBOSS in Sec. \ref{sec:results} where a comparison between PIP, PIP+ANG and CP correction schemea is presented on the eBOSS DR16 LSS catalogues.

\subsection{Effect of PIP and Angular Up-Weighting} \label{sec:ang_discussion}

\begin{figure}
    	\centering
		\includegraphics[scale=0.10]{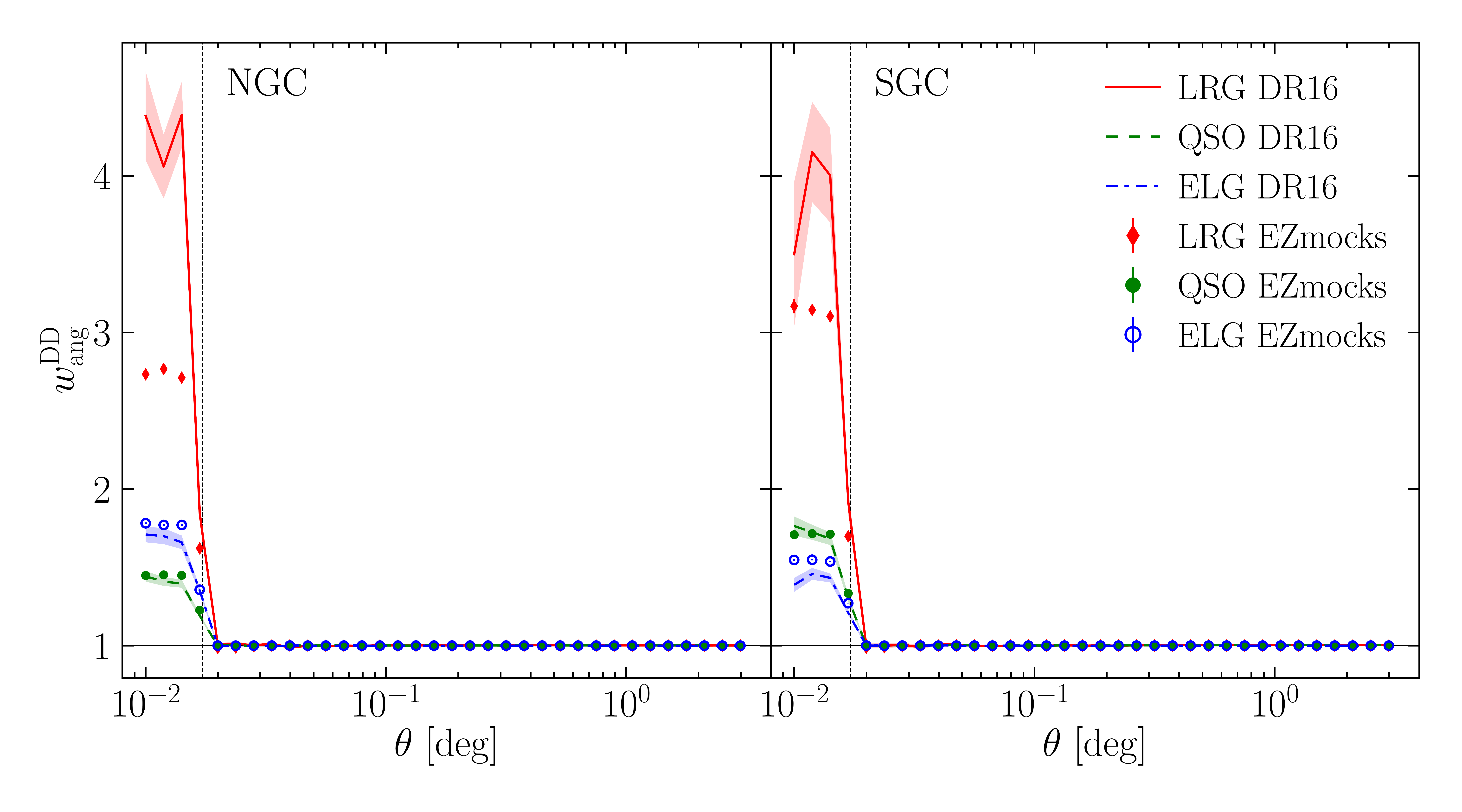}
		\caption{Angular weights used to up-weight $DD$ pair counts (eq. \ref{eq:pip+ang}) in the clustering measurements for luminous red galaxies (red), quasars (green) and emission-line galaxies (blue). Left and right panels display the weights for north (NGC) and south (SGC) galactic caps. Markers show the mean estimate from 100 EZmocks with error-bars being the error on the mean. Lines show the counterparts from DR16 catalogues with shaded bands showing the related errors on a single realisation obtained from 100 EZmocks.}\label{fig:w_ang}
\end{figure}
\begin{figure}
    	\centering
		\includegraphics[scale=0.095]{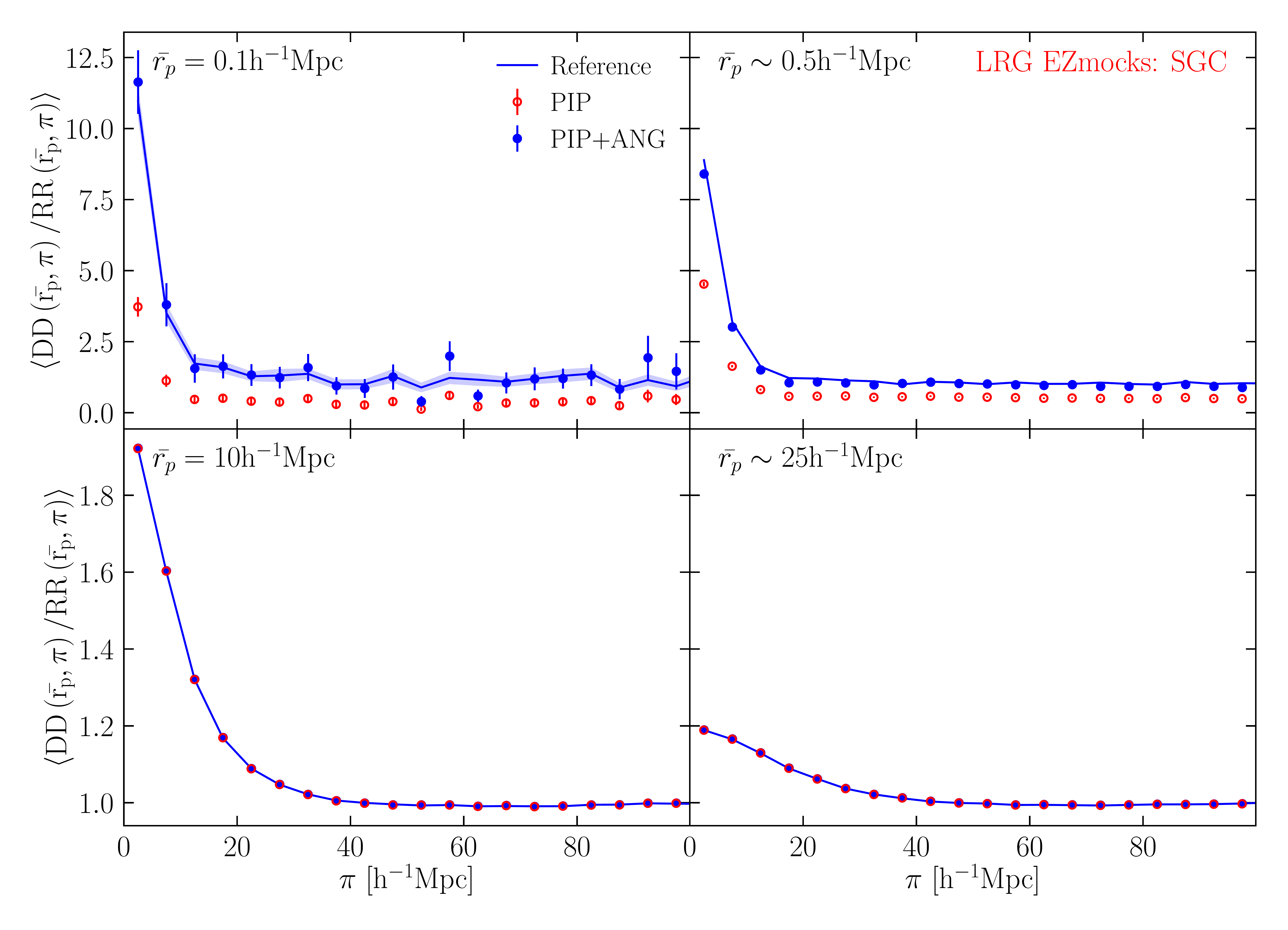}
		\caption{Ratio between $DD$ and $RR$ pair counts, averaged over 100 LRG EZmocks in SGC. Each panel displays the ratio as a function of the radial component $\pi$ of the pair separation, in a single bin in the transverse to the line-of-sight component $r_p$ of the pair separation. The $DD/RR$ ratio at two $r_p$ scales below the fibre collision scale is shown in the top panels while the same quantity for two values of $r_p$ larger than the fibre collision scale are plotted in the bottom panels. Blue lines show the mean estimate from 100 parent mocks while blue filled and red empty markers show the result of PIP and PIP+ANG correction applied to the corresponding mocks affected by fibre collisions. Shaded bands and error bars show the errors on the mean estimates.}\label{fig:ratio_dd-rr}
\end{figure}

We discuss here the impact that the PIP and angular up-weighting are expected to have on clustering measurements presented in Sec. \ref{sec:tests} and Sec. \ref{sec:results} for mocks and eBOSS DR16 catalogues, respectively.

The PIP up-weighting provides un-biased pair counts only if there are no pairs with zero probability of being targeted in a random realisation of the survey. In eBOSS, zero-probability pairs do originate from fibre collisions in single-pass regions. These zero-probability pairs however, are confined at angular separations below the fibre-collision angle $\theta^{\mathrm{fc}}=62\arcsec$. At these separations the PIP weighting properly up-weights pairs in the overlaps between multiple tiles but misses out those in areas covered by a single tile. At angular separations $\theta<\theta^{\mathrm{fc}}$ we therefore expect the PIP up-weighting to under-estimate the pair counts inferred from the spectroscopic sample with respect to those from the parent catalogue. PIP up-weighting provide virtually un-biased pair count at separations $\theta>\theta^{\mathrm{fc}}$ where no or a very small number of zero-probability pairs (those resulting from an optimisation of fibre assignment in collision groups with more than two objects) are expected. The effect is quantified by means of the angular weights $\wdd$ in eq. \eqref{eq:pip+ang} used to correct the $DD$ pair counts and shown in Fig. \ref{fig:w_ang} for luminous red galaxies (red), quasars (green) and emission-line galaxies (blue). Indeed, in all cases $\wdd$ is greater than unity for separations below $\theta^{\mathrm{fc}}$ and sharply reaches 1 for larger values of angular separation $\theta$. A difference is noticeable between NGC (left panel in Fig. \ref{fig:w_ang}) and SGC (right panel in Fig. \ref{fig:w_ang}) due to the different tiling density between the two caps. For LRGs and QSOs for example, tiles in the SEQUELS chunks in the NGC are more tightly packed with respect to those in eBOSS chunks. This increases the fraction of the area covered by more than one tile decreasing the fraction of zero-probability pairs, which in turn decreases $\wdd$. The angular weights $\wdd$ for LRGs tend to have significantly lower amplitudes in the EZmocks (markers with error-bars) compared to the eBOSS DR16 catalogue. This results from a lower intrinsic clustering of targets in the mocks with respect to eBOSS data combined with the fact that LRGs are targeted at the lowest priority. For quasars and emission-line galaxies the difference between mock and data in Fig \ref{fig:w_ang} is significantly smaller.

In order to assess how the PIP and PIP+ANG corrections impact the measurements of the projected correlation function or the multipole moments $\xi^{(\ell)}$ at different scales below the fibre collision scale, it is useful to work with the anisotropic two-point correlation function $\xirppi$, measured as a function of the parallel $\pi$ and transverse to the line-of-sight $r_p$ components of the pair separation, in terms of its natural estimator,
\begin{equation}
    \xi(\mathbf{s}) = \frac{DD(\mathbf{s})}{RR(\mathbf{s})}-1. \label{eq:Nt}
\end{equation}
With respect to the angular separations, the transverse scale that corresponds to the fibre-collision angle $\theta^{\mathrm{fc}}$ in the fiducial cosmology varies with redshift. We denote with $r_p^{\mathrm{fc}}$ the transverse scale spanned by $\theta^{\rm{fc}}$ at the maximum redshift of the sample. The PIP-corrected $DD(r_p,\pi)$ pair counts are then expected to be negatively biased at any $\pi$ for $r_p<r_p^{\rm{fc}}$ that in turn results in an under-estimation of the ratio $DD/RR$ and thus of the two-point correlation function with respect to the reference one. At scales $r_p>r_p^{\rm{fc}}$ all pairs are at angular separations $\theta>\theta^{\rm{fc}}$, a regime where PIP-corrections are un-biased, we expect $DD(r_p,\pi)$ corrected using PIP weights, and thus also the anisotropic two-point correlation function $\xirppi$, to match its value from the reference parent sample. This is illustrated in Fig. \ref{fig:ratio_dd-rr} for luminous red galaxies using 100 EZmocks where we can compare the PIP up-weighted measurements with the ones from parent samples. In Fig. \ref{fig:ratio_dd-rr} we show the ratio $DD(r_p,\pi)/RR(r_p,\pi)$ from eq. \eqref{eq:Nt} at two transverse scales smaller (top panels) and two larger than $r_p^{\rm{fc}}\sim0.7\mhmpc$. As expected, the PIP correction strongly underestimate the $DD/RR$ ratio, or equivalently the $DD$ pair counts, for $\bar{r_p}<0.7\mhmpc$ where the angular up-weighting is needed to properly account for zero-probability pairs. At transverse scales $\bar{r_p}>0.7\mhmpc$, on the other hand, angular up-weighting has negligible effect as $\wdd\sim1$ and the PIP and PIP+ANG up-weighting are both un-biased.

\subsection{Projected Correlation Function} \label{sec:wp_mocks}
 \begin{figure}
    	\centering
		\includegraphics[scale=0.09]{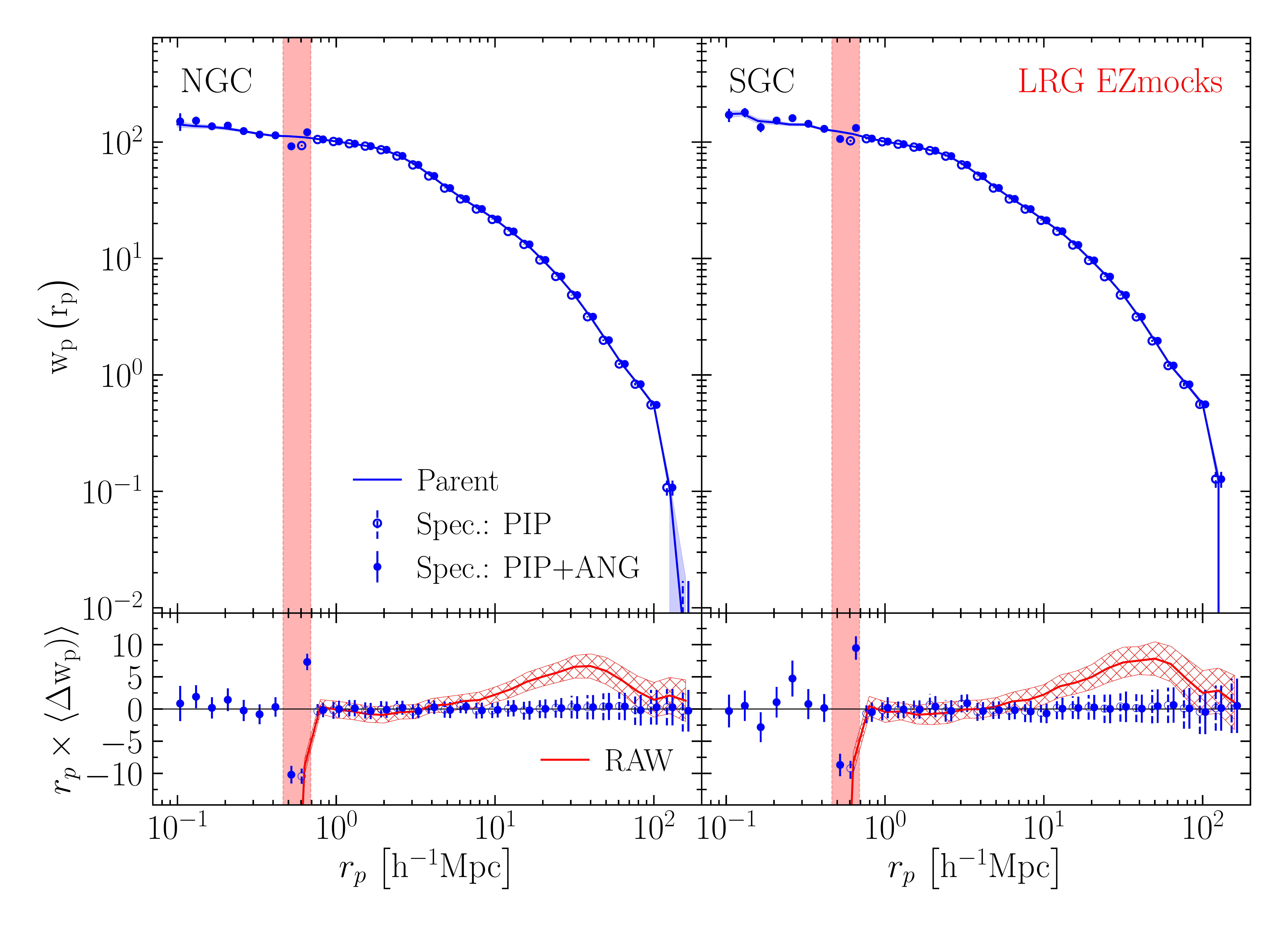}
		\caption{Projected correlation function of LRG EZmocks built as described in Sec. \ref{sec:tests} in the two caps, NGC (left panels) and SGC (right panels). Top panels: mean estimate from 100 parent mocks (continuous lines), from catalogues affected by fibre collisions corrected using the PIP-only (empty markers with dashed error-bars) and corrected using the combined PIP and angular up-weighting (PIP+ANG, filled points with continuous error-bars). The blue shaded bands and error bars show the error on the mean. The vertical red shaded bands show the transverse scales in the EZmocks fiducial cosmology corresponding to the fibre-collision angle between the minimum and maximum redshift of the sample. For separations $r_p$ larger than the fibre-collision scale PIP-only and the joint PIP+ANG corrections provide almost identical results. Therefore, empty (PIP) and filled (PIP+ANG) markers are these scales can not be easily distinguished.	Empty markers at scales smaller than the fibre-collision are not visible in the plot because they are well below the minimum limit set on the y axis. Bottom panels: mean of the differences between the corrected measurements from mocks affected by fibre collisions and the corresponding parent mock. To reduce the range of variation, each quantity in the bottom panel is multiplied by $r_p$. For comparison, in the bottom panels that show the differences, the red continuous lines and hatch regions (RAW) show the mean measurements from spectroscopic mocks and related errors in the case where no correction for fibre collisions is applied.}\label{fig:wp_lrg_mocks}
	\end{figure}

 \begin{figure}
    	\centering
		\includegraphics[scale=0.10]{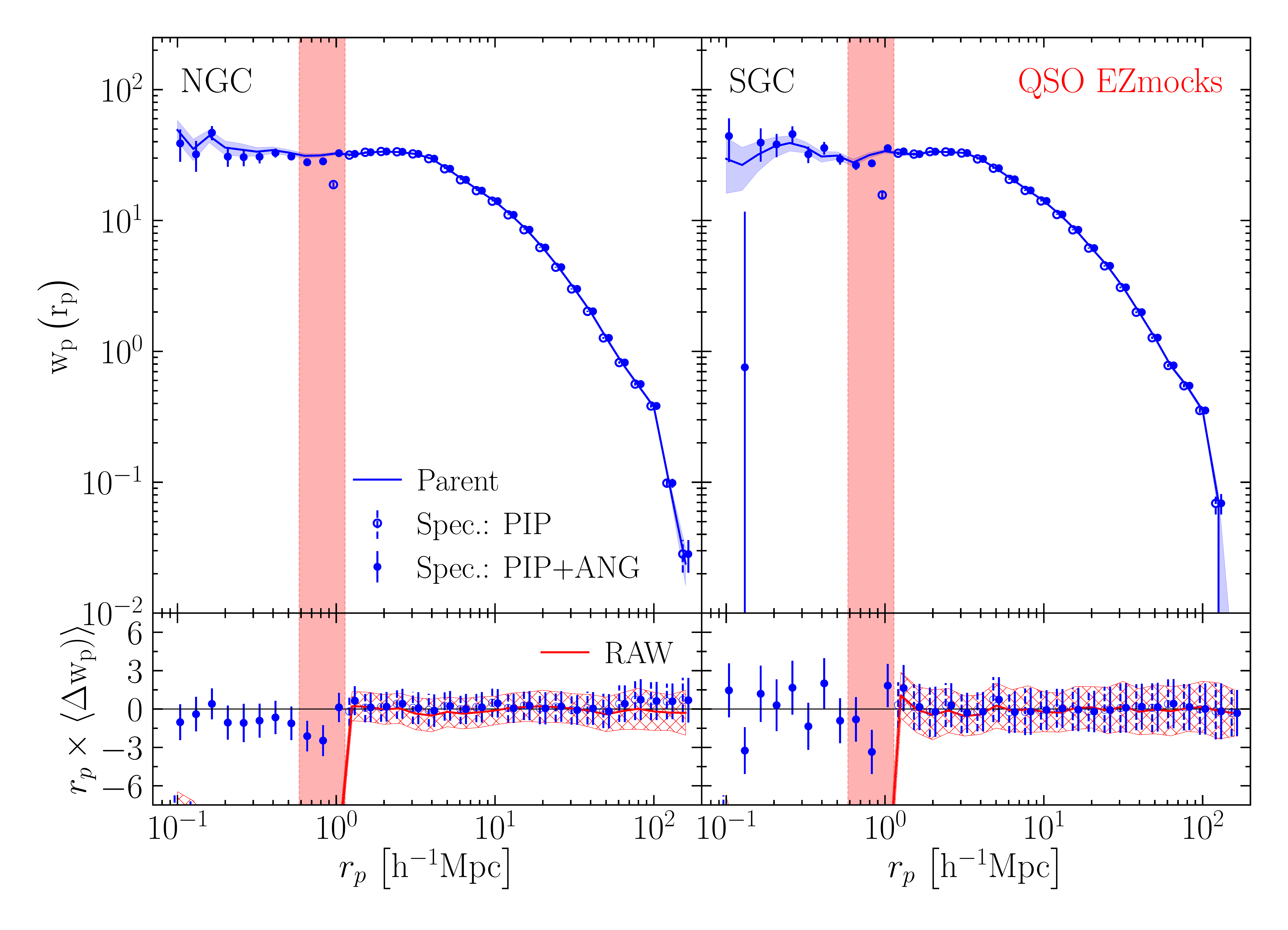}
		\caption{Same as in Fig. \ref{fig:wp_lrg_mocks} here for the QSO EZmocks.}\label{fig:wp_qso_mocks}
	\end{figure}

 \begin{figure}
    	\centering
		\includegraphics[scale=0.10]{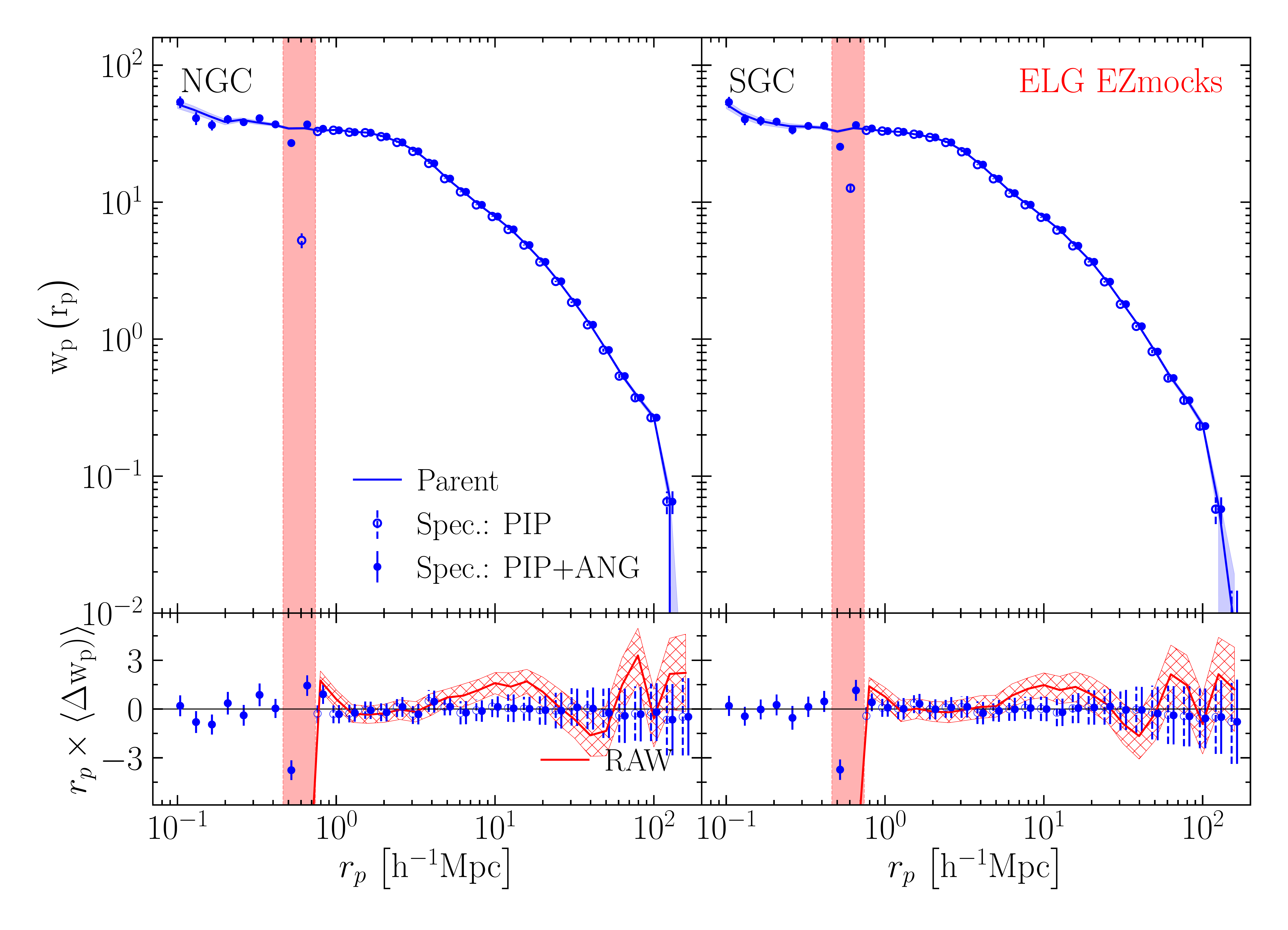}
		\caption{Same as in Fig. \ref{fig:wp_lrg_mocks} here for the ELG EZmocks.}\label{fig:wp_elg_mocks}
	\end{figure}

Mean estimates of the projected correlation function $\mwp$ and the corresponding statistical errors from a set of 100 mocks are shown in the top panels of Fig. \ref{fig:wp_lrg_mocks}-\ref{fig:wp_elg_mocks} for the three tracers. The bottom panels of Fig. \ref{fig:wp_lrg_mocks}-\ref{fig:wp_elg_mocks} show the difference between measurements from the spectroscopic (see Sec. \ref{sec:contaminants}) and the parent mocks.

In the bottom panels of Fig. \ref{fig:wp_lrg_mocks}-\ref{fig:wp_elg_mocks} we report the case where no correction for fibre collisions or survey completeness is applied (red continuous lines with hatch error regions) to the measurement from the spectroscopic mocks. Given the sparsity of the sample and being targeted at higher priority, missing observations have negligible effect on the clustering of quasars (see Fig. \ref{fig:wp_qso_mocks}) on scales larger than the fibre-collision scale where missing targets resemble the effect of a random depletion. Raw estimate of the correlation function of luminous red galaxies (see Fig. \ref{fig:wp_lrg_mocks}), on the other hand, shows negligible offset with respect to the reference one on scales $\sim1$-$10\mhmpc$ and scales $\gtrsim100\mhmpc$, while an offset at $\sim2\sigma$ level is clearly visible on scales between $\sim30$-$100\mhmpc$. Raw measurements of the projected correlation function of emission-line galaxies mocks in Fig. \ref{fig:wp_elg_mocks} show an overall agreement with the clustering of the parent samples with a marginal offset at scales of $\sim10\mhmpc$. These raw measurements show that although the fibre collisions have a limited impact compared to the statistical errors, its strength strongly depends on the intrinsic clustering of the tracers and the features of the particular fibre-assignment algorithm adopted.

Measurements from spectroscopic mocks, corrected using the PIP technique and averaged over 100 mocks, are shown as empty markers in the top panels of Fig. \ref{fig:wp_lrg_mocks}-\ref{fig:wp_elg_mocks} along with the reference measurements from parent mocks (continuous lines). For all three tracers the agreement between the PIP-only corrected and reference measurements is remarkable for transverse scales larger than $\sim 1\mhmpc$ (bottom panels in Fig. \ref{fig:wp_lrg_mocks}-\ref{fig:wp_elg_mocks}). However, PIP weighting fails to recover the input clustering at transverse scales $r_p$ below the fibre-collision scale (vertical red shaded bands in Fig. \ref{fig:wp_lrg_mocks}-\ref{fig:wp_elg_mocks}). The interpretation of the systematic offset at these scales, when PIP-only up-weighting is applied, follows from the discussion in Sec. \ref{sec:ang_discussion}. In particular, in the limit of small $r_p$ and small $\pi$, where targets are strongly clustered, the anisotropic correlation function $\xirppi$ is well approximated by $DD/RR$ and the PIP-corrected $\xirppi$ results are offset by a factor of $\wdd$ with respect to the reference. At small $r_p$ and large $\pi$, specifically between $\pi\sim50\mhmpc$ and the upper limit in the integral in eq. \eqref{eq:wp} fixed at $80\mhmpc$, i.e. in the regime of weak intrinsic clustering, $DD/RR$ approaches unity and $\xirppi$ tends to 0. At these scales, the underestimation in the PIP up-weighted $DD$ pair counts reduces the corresponding $\xirppi$ to very small or negative values and the scaling by $\wdd$ between the PIP-corrected and reference $\xirppi$ is not valid anymore. This drives the PIP-corrected projected correlation function $\mwp$, obtained integrating $\xirppi$ in eq. \eqref{eq:wp}, to values below the lower limit on y-axis shown in the top panels of Fig. \ref{fig:wp_lrg_mocks}-\ref{fig:wp_elg_mocks} enhancing the relative difference between PIP up-weighted and reference measurements well above the factor of $\wdd$.

The angular up-weighting uses the fraction of close pairs lost in single-pass regions to restore the $DD/RR$ ratio at small transverse scales $r_p$, to its expectation values. As a result, when PIP weights are supplemented with the angular up-weighting (filled markers with error-bars in Fig. \ref{fig:wp_lrg_mocks}-\ref{fig:wp_elg_mocks}), we are able to successfully recover the clustering signal down to very small scales $\sim0.1\mhmpc$ without altering the large-scale measurements.

It is important to stress that, as opposed to the raw measurements, the performance of both the PIP and joint PIP and angular up-weighting does not vary with the type of tracers. As anticipated in Sec. \ref{sec:survey_realizations}, the optimization performed by the fibre assignment algorithm within the collision groups can give rise to the `zero-probability' pairs at the scale of fibre collisions. This is likely to be the source of the small deviation in the PIP+ANG corrected measurements at the close-pair scales (vertical red shaded bands) seen in Fig. \ref{fig:wp_lrg_mocks}-\ref{fig:wp_elg_mocks} with respect to the reference. The effect, stronger for the LRGs in Fig. \ref{fig:wp_lrg_mocks}, is well within the statistical error for a single realisation and becomes evident only when averaged over a high number of samples. We therefore do not consider this further: as demonstrated in the plot it is small and limited to a narrow range of scales close to the collision scale.

\subsection{Multipoles}\label{sec:mps_mocks}

 \begin{figure}
    	\centering
		\includegraphics[scale=0.10]{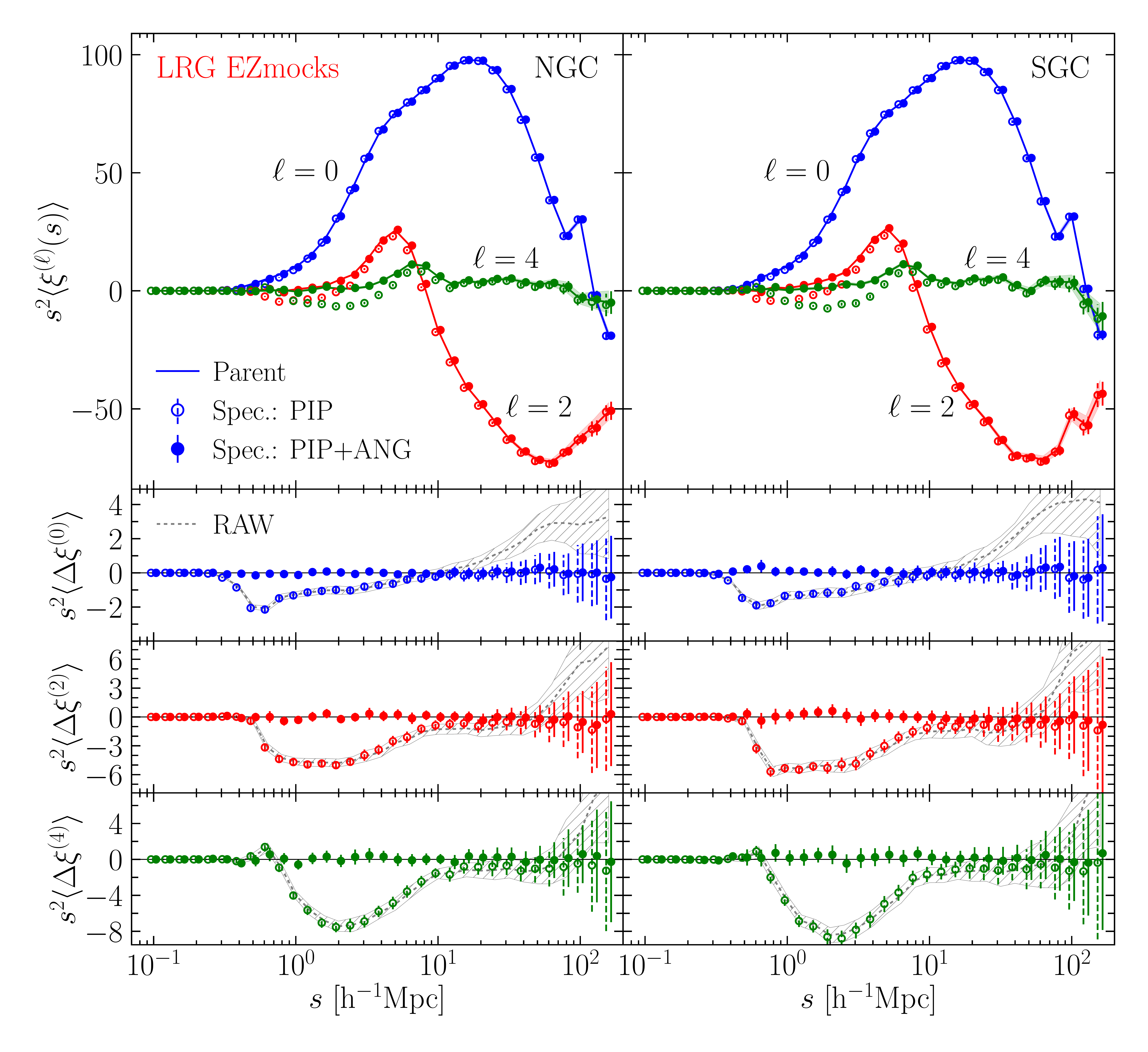}
		\caption{Multipoles of the redshift-space two-point correlation functions from 100 EZmocks of luminous red galaxies. In all panels, measurements are averaged over 100 mocks and the error-bars correspond to the error on the mean. Top row: mean measurements from the reference parent mock catalogues (lines with shaded bands) and from the corresponding samples of targeted objects that are corrected using PIP-only (empty markers with dashed error-bars) and combined PIP and angular (PIP+ANG, filled markers with thick error-bars) weighting schemes. Bottom three panels: mean difference between the raw measurements (grey dotted lines and hatches), PIP (empty markers), PIP+ANG (filled markers) corrections and the reference measurements.}\label{fig:mps_lrg_mocks}
	\end{figure}
 \begin{figure}
    	\centering
		\includegraphics[scale=0.10]{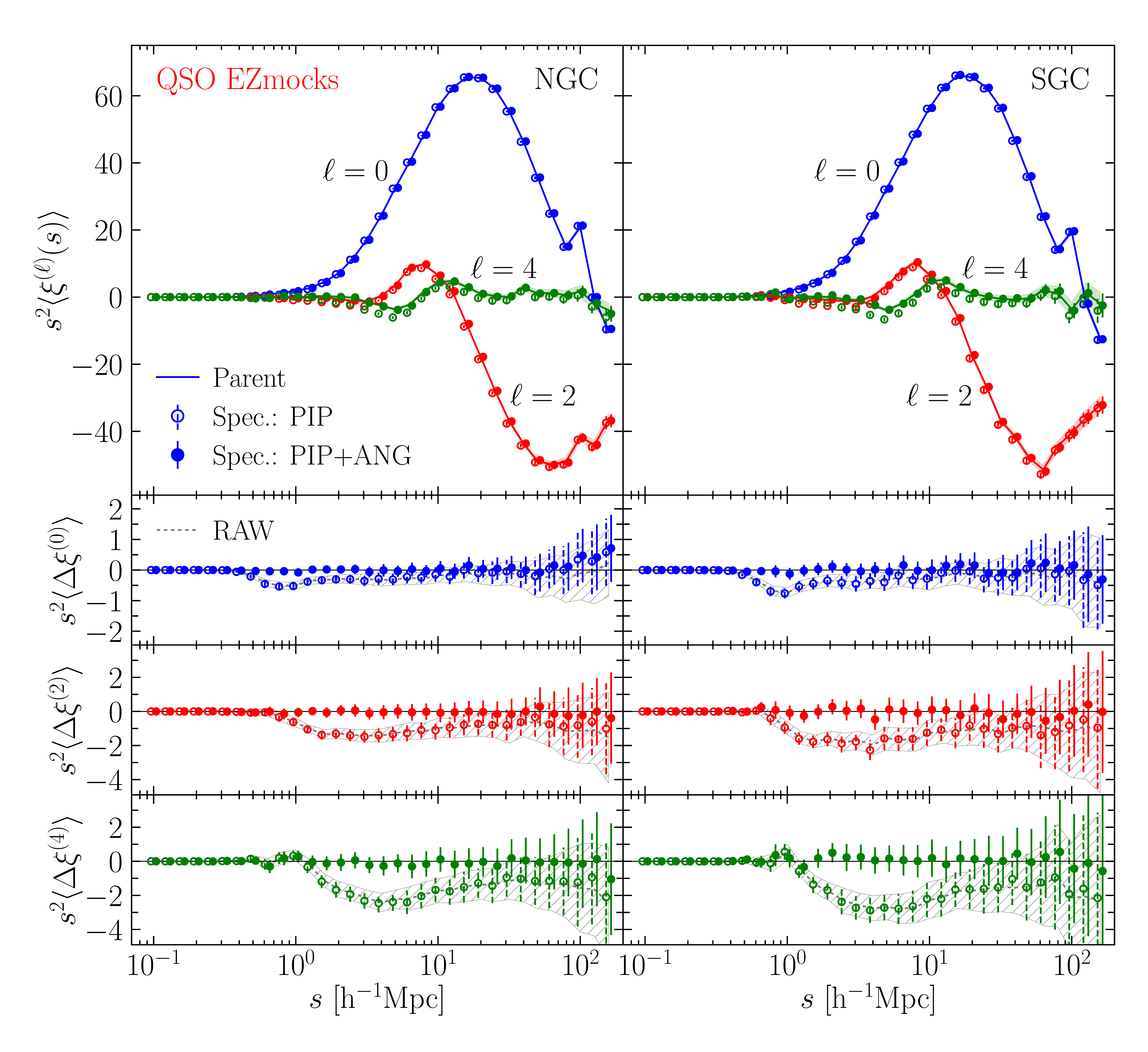}
		\caption{Same as in Fig. \ref{fig:mps_lrg_mocks}, here for the quasar mock samples.}\label{fig:mps_qso_mocks}
	\end{figure}
 \begin{figure}
    	\centering
		\includegraphics[scale=0.10]{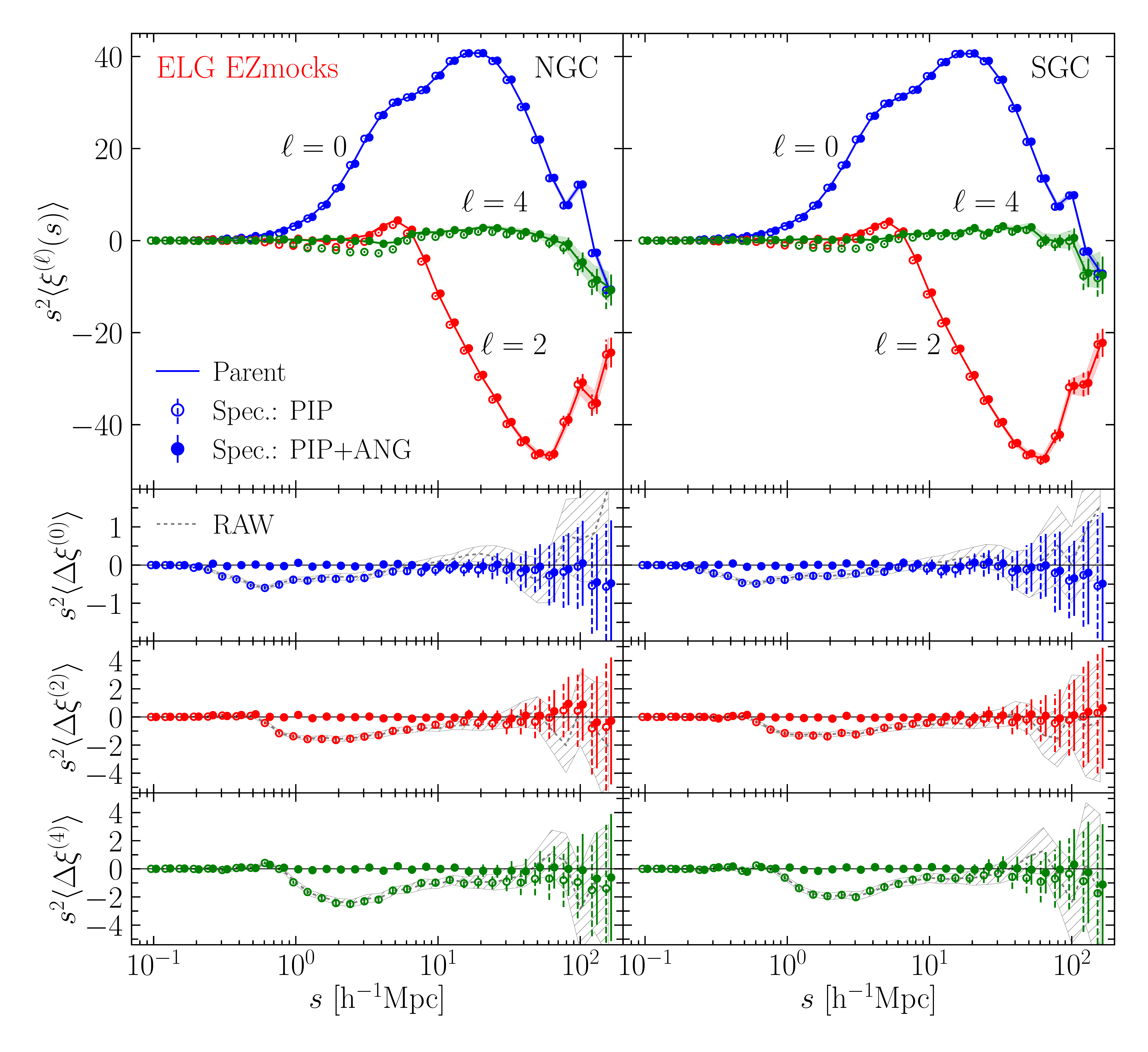}
		\caption{Same as in Fig. \ref{fig:mps_lrg_mocks}, here for the mock samples of emission-line galaxies.}\label{fig:mps_elg_mocks}
	\end{figure}

We now test the corrections for the multipoles of the redshift-space two-point correlation function. We limit the tests to the first three even multipoles, namely the monopole, quadrupole and hexadecapole that are mostly used to detect and model the redshift-space distortions and the baryon acoustic oscillations in large galaxy redshift surveys. Results are shown in Fig. \ref{fig:mps_lrg_mocks}-\ref{fig:mps_elg_mocks}, and match the results presented for the projected correlation functions in Fig. \ref{fig:wp_lrg_mocks}-\ref{fig:wp_elg_mocks}.

Differences between `raw' measurements from spectroscopic mocks not corrected for fibre collisions or survey completeness, and the reference measurements from the parent mocks are shown in the bottom panels of Fig. \ref{fig:mps_lrg_mocks}-\ref{fig:mps_elg_mocks} (dashed lines with hatch error regions). Raw measurements tend to under-estimate the reference clustering up to scales of $\sim10$-$20\mhmpc$ for all tracers although the bias is more severe for luminous red galaxies due to their higher clustering and being targeted at the lowest priority. At larger scales a mild bias appears for luminous red galaxies while it is not clearly visible for quasars and emission-line galaxies due to relatively large statistical errors at these scales.

The PIP corrections are un-biased on scales larger than $\sim10\rm{-}20\mhmpc$ recovering the input clustering signal within less than $\sim1$-$\sigma$ errors.  However, they systematically under-estimate the clustering at smaller scales. The nature of the systematic bias in the PIP-corrected measurements of the multipole moments $\xi^{(\ell)}$ is the same discussed in Sec. \ref{sec:ang_discussion} and in Sec. \ref{sec:wp_mocks} for the projected correlation function. However, the effect, confined at small transverse scales $r_p$ in the anisotropic $\xirppi$ and projected $\mwp$ correlation functions, is now spread to all angle-averaged separations $s$ in the multipoles $\xi^{(\ell)}$. At angle-averaged scales smaller than the transverse fibre-collision scale $s<r_p^{\rm{fc}}$, PIP weighting underestimates the reference $DD(s,\mu)$ pair count at any value of $\mu$ resulting in a strong negative bias in the measured multipoles. At scales $s\gtrsim r_p^{\rm{fc}}$ the bias is reduced due to the fact that the PIP-corrected $DD(s,\mu)$ pair counts are underestimated, with respect to the reference, only between $1<\mu<\mu^{\rm{fc}}$ with,
    \begin{equation}
        \mu^{\rm{fc}} = \left[1-\left({r_p^{\rm{fc}}}/{s}\right)\right]^{1/2}. \label{eq:mu_fc}
    \end{equation}
At scales $s$ significantly larger than $r_p^{\rm{fc}}$, $\mu^{\rm{fc}}$ approaches unity and the underestimate in the $DD$ pair counts (see Sec. \ref{sec:ang_discussion}) between $[\mu^{\rm{fc}},1]$ has negligible effect on the multipole moments $\xi^{(\ell)}$. The systematic bias at a given scale $s$ is also higher for higher order multipoles. This follows from the $\mu$ dependence of the Legendre polynomials in eq. \eqref{eq:mps}.

As for the projected correlation function $\mwp$, the angular weights properly up-weight the $DD$ pair counts below the fibre collision scale providing un-biased measurements of the anisotropic correlation function $\xi(s,\mu)$ and its multipoles. The combined PIP+ANG correction results are un-biased at all scales explored in this work at a level well below the statistical errors, as shown in the bottom panels of Fig. \ref{fig:mps_lrg_mocks}-\ref{fig:mps_elg_mocks}.

\section{Results}\label{sec:results}
The tests performed using the EZmock catalogues in Sec. \ref{sec:tests} showed that we can successfully recover the input clustering signal down to very small scales, well within the 1-halo term. We now apply the same corrections to the eBOSS DR16 LSS catalogues. Since we deal with a single catalogue for each tracer, we increase the number of survey realisations, used to infer the selection probabilities, from 310 used for each mock catalogue to 1860. The implementation of the angular up-weighting for the DR16 catalogues is slightly different from that outlined in Sec. \ref{sec:tests} for the mock catalogues. In particular, for each tracer we now use the corresponding full input sample as the parent catalogue to compute the quantities labelled with -par, and their targeted sub-samples to compute the ones denoted with -fib in eq. \eqref{eq:pip+ang} regardless of their redshifts. We also compare the novel correction schemes to the standard CP weighting used in eBOSS DR16 cosmology papers.

The shaded bands in Fig. \ref{fig:wp_lrg_data}-\ref{fig:mps_elg_data} show the statistical error on a single eBOSS realization derived using a set of 100 EZmocks. These mocks only provide approximate realisations of the galaxy distribution on small scales, and so we do not expect the error-bars to fully capture the fluctuations observed in eBOSS data at scales $\lesssim 1\mhmpc$.

\subsection{Projected Correlation Function}
 \begin{figure}
    	\centering
		\includegraphics[scale=0.10]{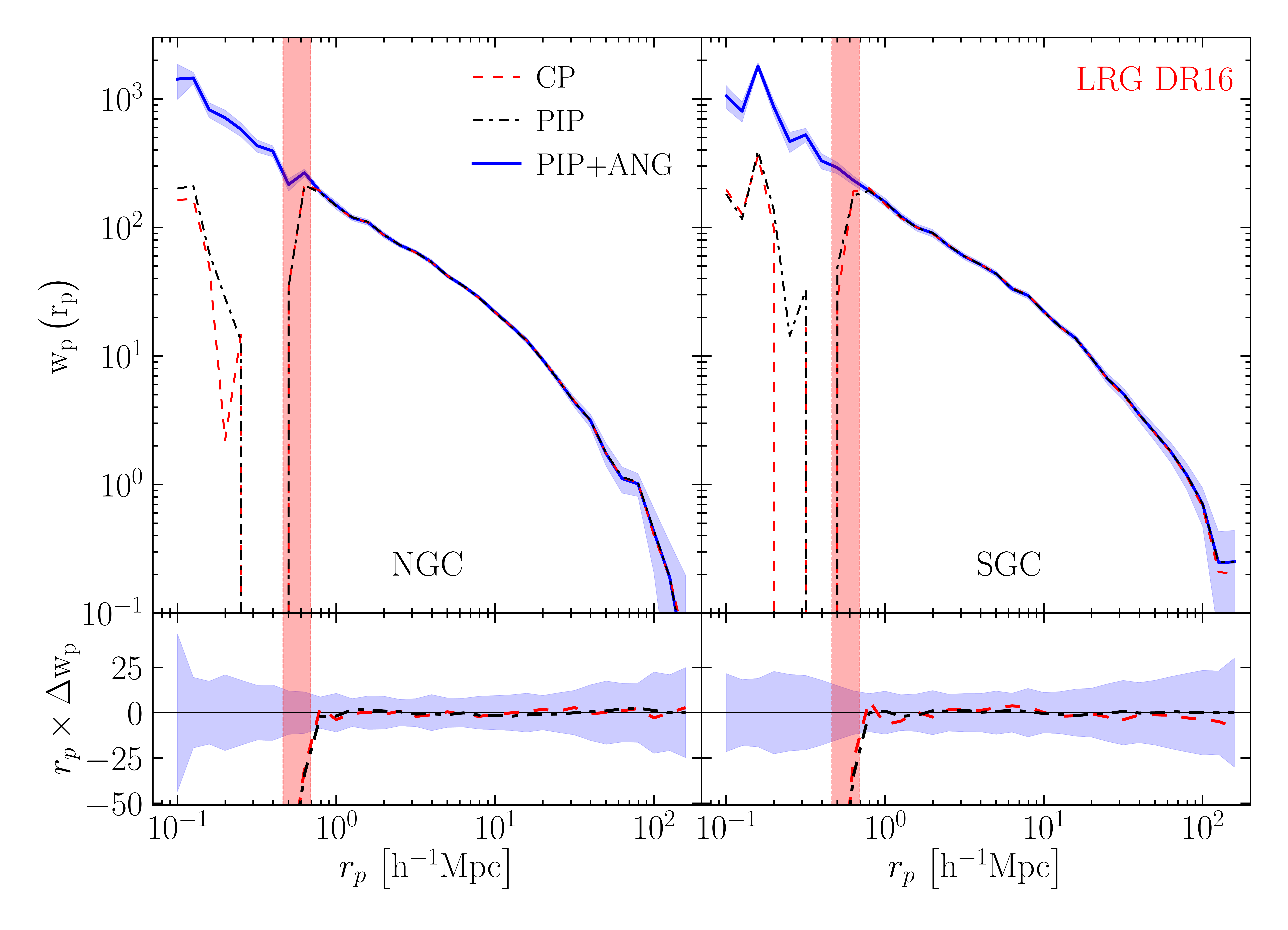}
		\caption{Top panel: measurements of the projected correlation function of the eBOSS DR16 LRG sample using different correction schemes: close-pair weighting (CP, dashed red lines), PIP weighting (PIP, black dashed-dotted line), combined PIP and angular up-weighting (PIP+ANG, blue continuous lines). The shaded band shows the 1-$\sigma$ errors estimated using 100 mock samples. Bottom panel: difference between the measurements obtained using CP and PIP weighting with respect to measurements using combined PIP and angular up-weighting. The shaded band shows the 1-$\sigma$ statistical error. To reduce the range of variation, each quantity in the bottom panel is multiplied by $r_p$. As in Fig. \ref{fig:wp_lrg_mocks} the vertical shaded red bands show the transverse scales corresponding to the fibre-collision angle between the minimum and maximum redshift of the sample in the eBOSS fiducial cosmology.}\label{fig:wp_lrg_data}
	\end{figure}

 \begin{figure}
    	\centering
		\includegraphics[scale=0.10]{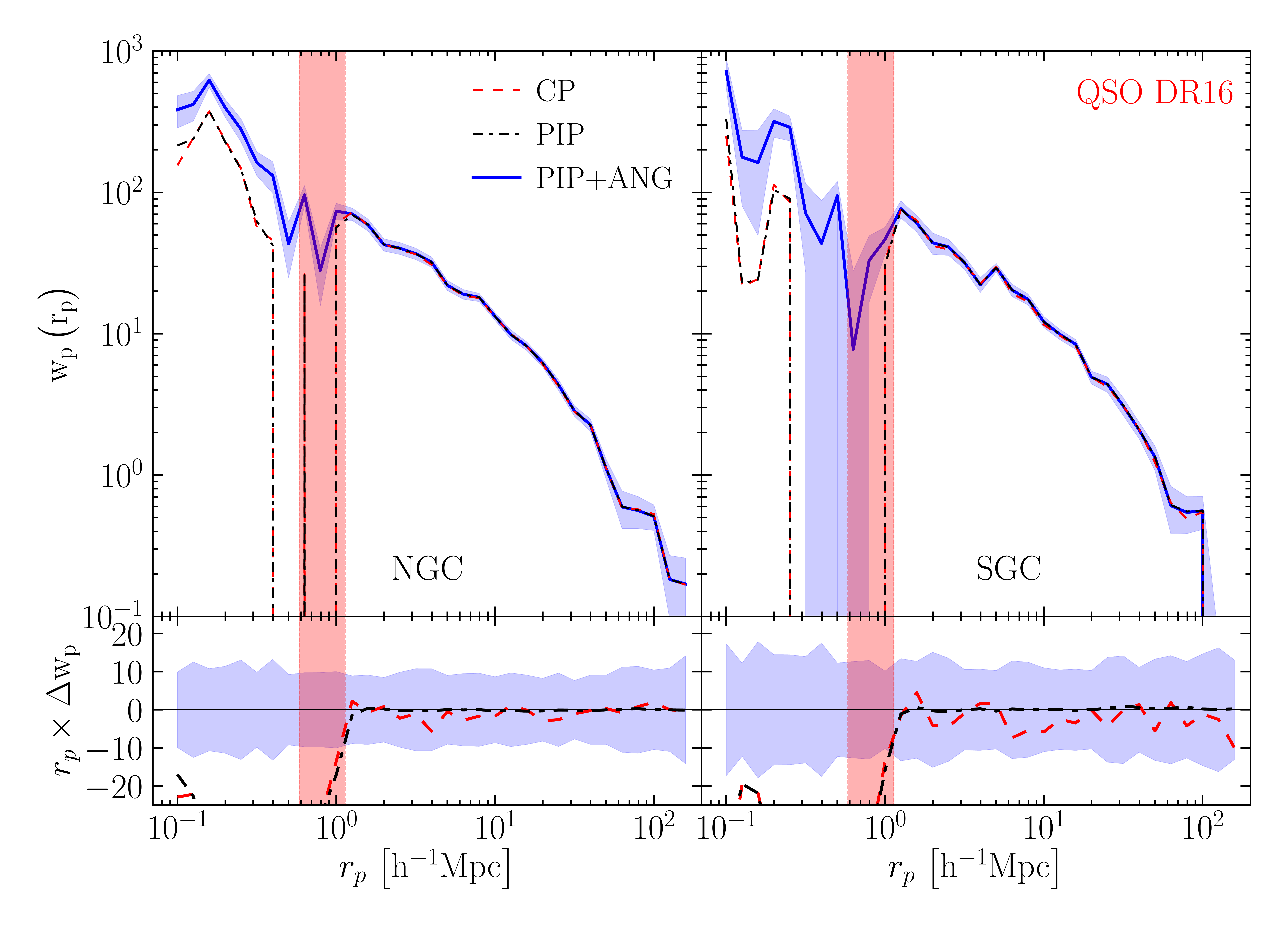}
		\caption{Equivalent of Fig. \ref{fig:wp_lrg_data} for the eBOSS DR16 quasar catalogue.}\label{fig:wp_qso_data}
\end{figure}

\begin{figure}
    	\centering
		\includegraphics[scale=0.10]{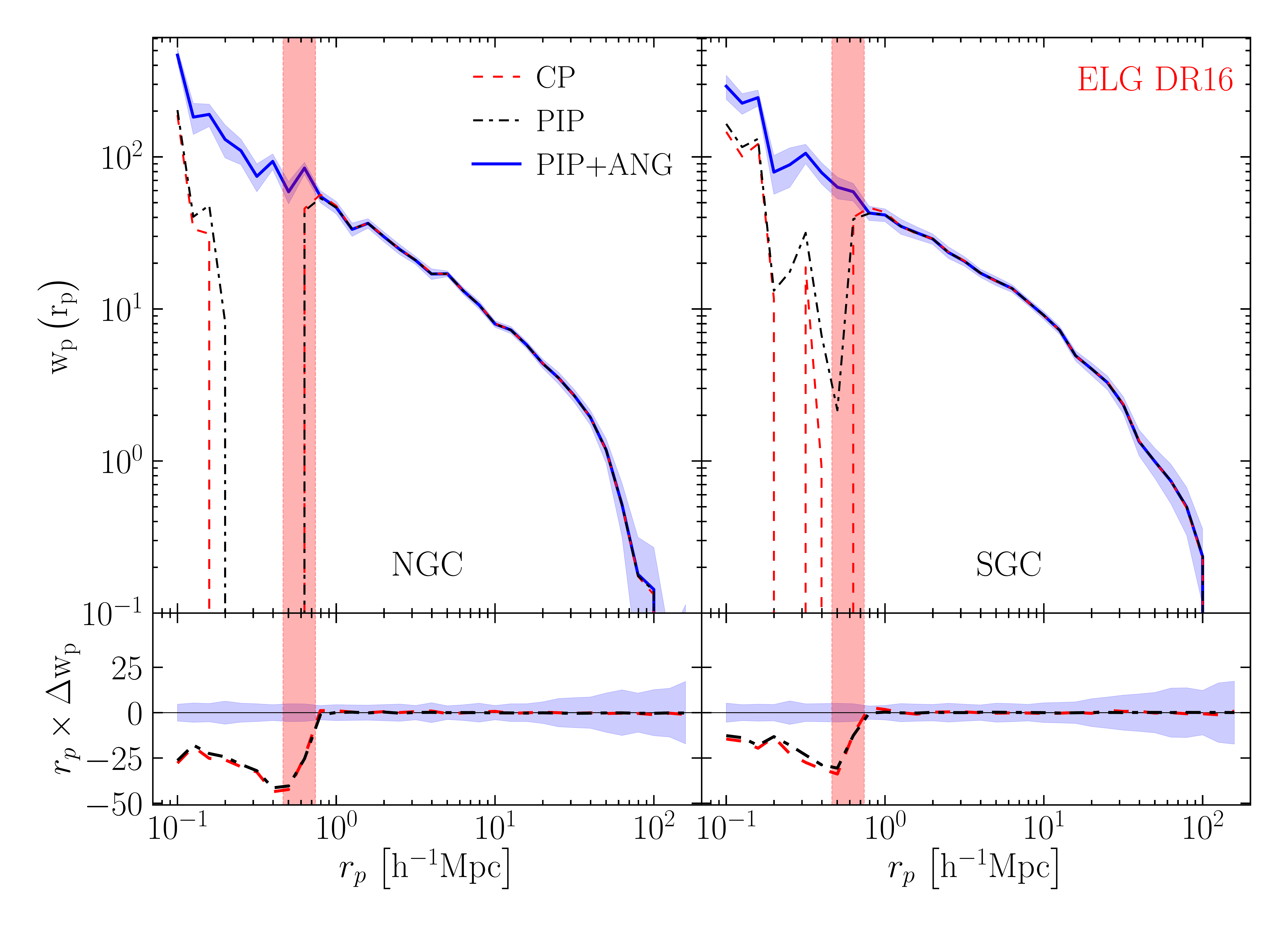}
		\caption{Equivalent of Fig. \ref{fig:wp_lrg_data} for the eBOSS DR16 catalogue of emission-line galaxies.}\label{fig:wp_elg_data}
\end{figure}

Figures \ref{fig:wp_lrg_data}, \ref{fig:wp_qso_data} and \ref{fig:wp_elg_data} show the measured projected correlation functions from the eBOSS DR16 samples of luminous red galaxies, quasars and emission-line galaxies respectively. Measurements (top panels) are corrected using the standard CP (red dashed lines), PIP (black dash-dotted lines) and the joint PIP and angular up-weighting (PIP+ANG, blue thick lines). In the bottom panels of the same figures we show the difference of the standard CP and PIP corrections with respect to the joint PIP+ANG up-weighting taken as the reference since it is found to be un-biased within data precision over all scales in the tests performed on mock catalogues.

The PIP up-weighting matches the PIP+ANG correction for scales larger than the fibre-collision scales (vertical red shaded bands) while it significantly underestimates the clustering at smaller scales. The CP corrections perform very similarly to the PIP ones with deviations consistent with the statistical noise on scales larger than the collision scale. The agreement between the CP and PIP corrections at scales larger than the fibre-collision scale shows that the selection probabilities are highly un-correlated on these scales and can be well approximated using an empirical prescription such as the nearest-neighbour method. This directly follows from the features of the eBOSS fibre assignment algorithm that uses a random seed to resolve collisions. The strong difference between the CP and PIP corrections with respect to the PIP+ANG weighting below the fibre collision scale (vertical red shaded bands in Fig. \ref{fig:wp_lrg_data}-\ref{fig:wp_elg_data}) is due to the effect discussed in Sec. \ref{sec:ang_discussion} and reflects the trend observed for mocks in Sec. \ref{sec:wp_mocks}. Comparing Fig. \ref{fig:wp_lrg_data}-\ref{fig:wp_elg_data} for eBOSS DR16 samples with their equivalent for mocks in Fig. \ref{fig:wp_lrg_mocks}-\ref{fig:wp_elg_mocks} it is clear that the eBOSS DR16 targets show a higher intrinsic clustering at scales below $~1\mhmpc$. Therefore, collisions are expected to occur at a higher rate in eBOSS catalogue with respect to the mocks. This is the source of the increase in the absolute size of the small-scale difference between PIP/CP and the PIP+ANG weighting scheme between the DR16 data and the mocks.

\subsection{Multipoles}
 \begin{figure}
    	\centering
		\includegraphics[scale=0.10]{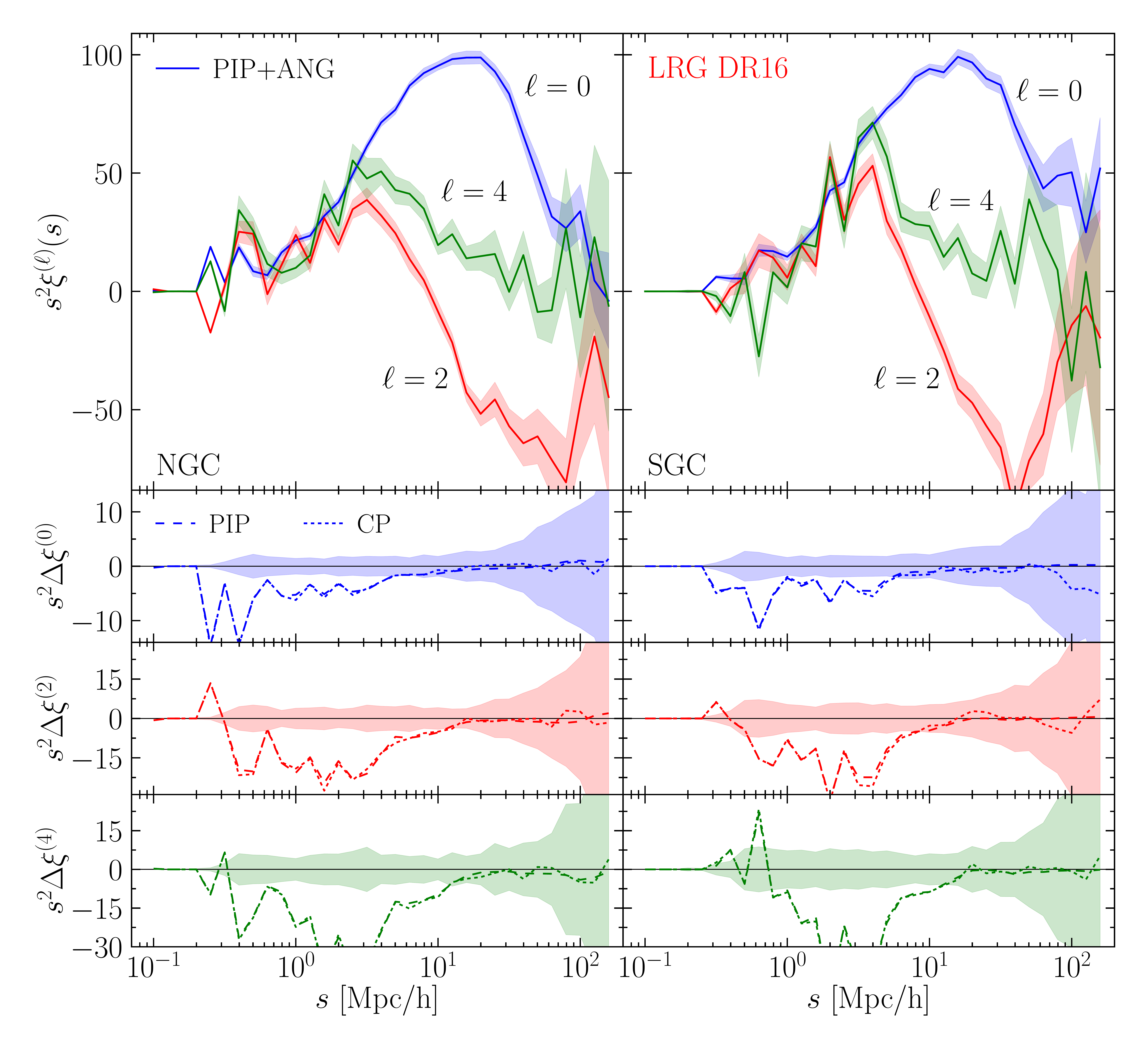}
		\caption{Top panel: measurements of the redshift-space multipole correlation functions of eBOSS DR16 LRG sample corrected using the combined PIP and angular up-weighting. The shaded band shows 1-$\sigma$ errors estimated using 100 mock samples. Bottom panels: difference of the measurements obtained using CP and PIP weighting with respect to the one using combined PIP and angular up-weighting. The shaded bands show the 1-$\sigma$ statistical error from the top panels.}\label{fig:mps_lrg_data}
	\end{figure}

 \begin{figure}
    	\centering
		\includegraphics[scale=0.10]{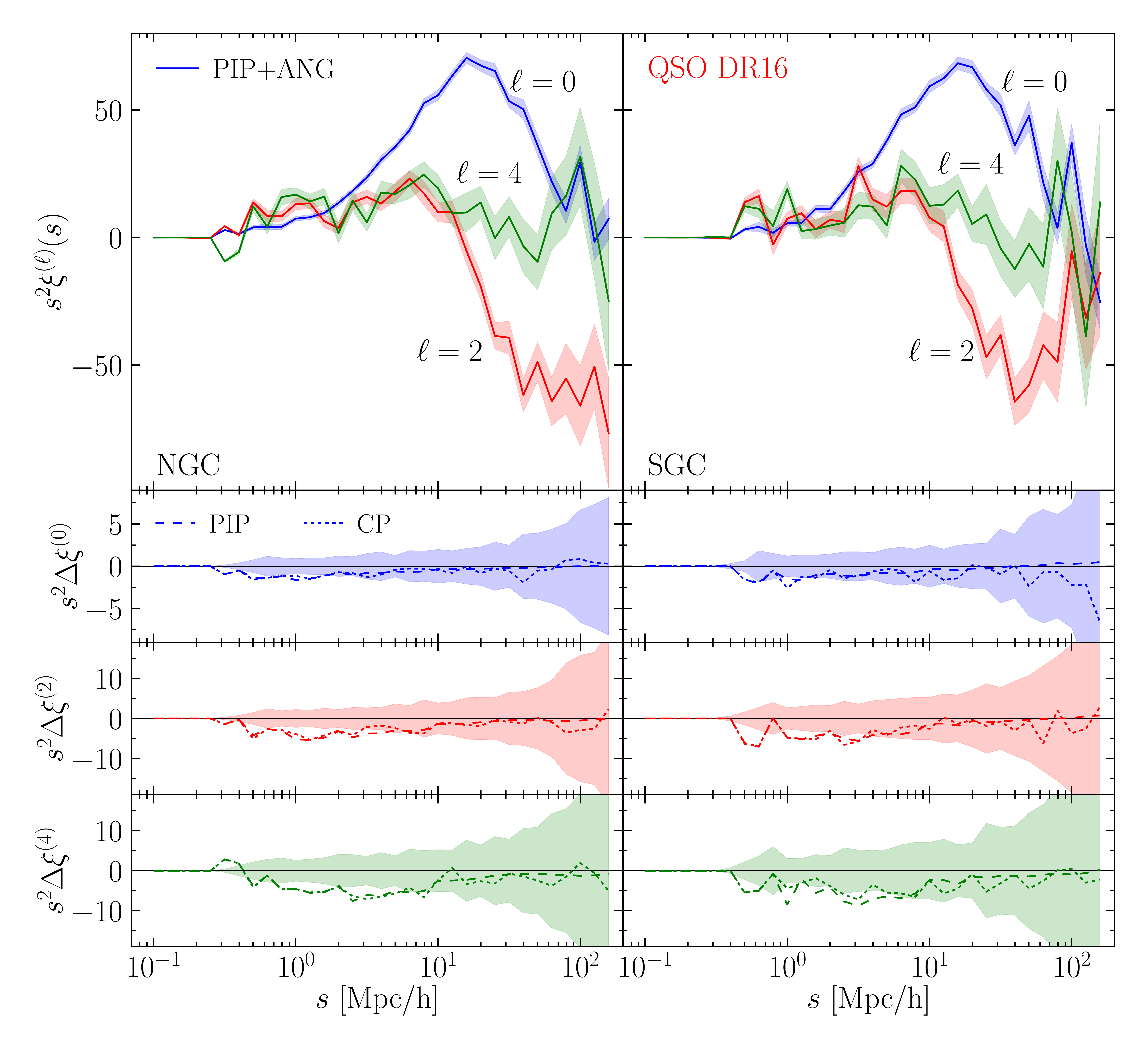}
		\caption{Same as in Fig. \ref{fig:mps_lrg_data}, here for the eBOSS DR16 quasar catalogue.}\label{fig:mps_qso_data}
	\end{figure}

 \begin{figure}
    	\centering
		\includegraphics[scale=0.10]{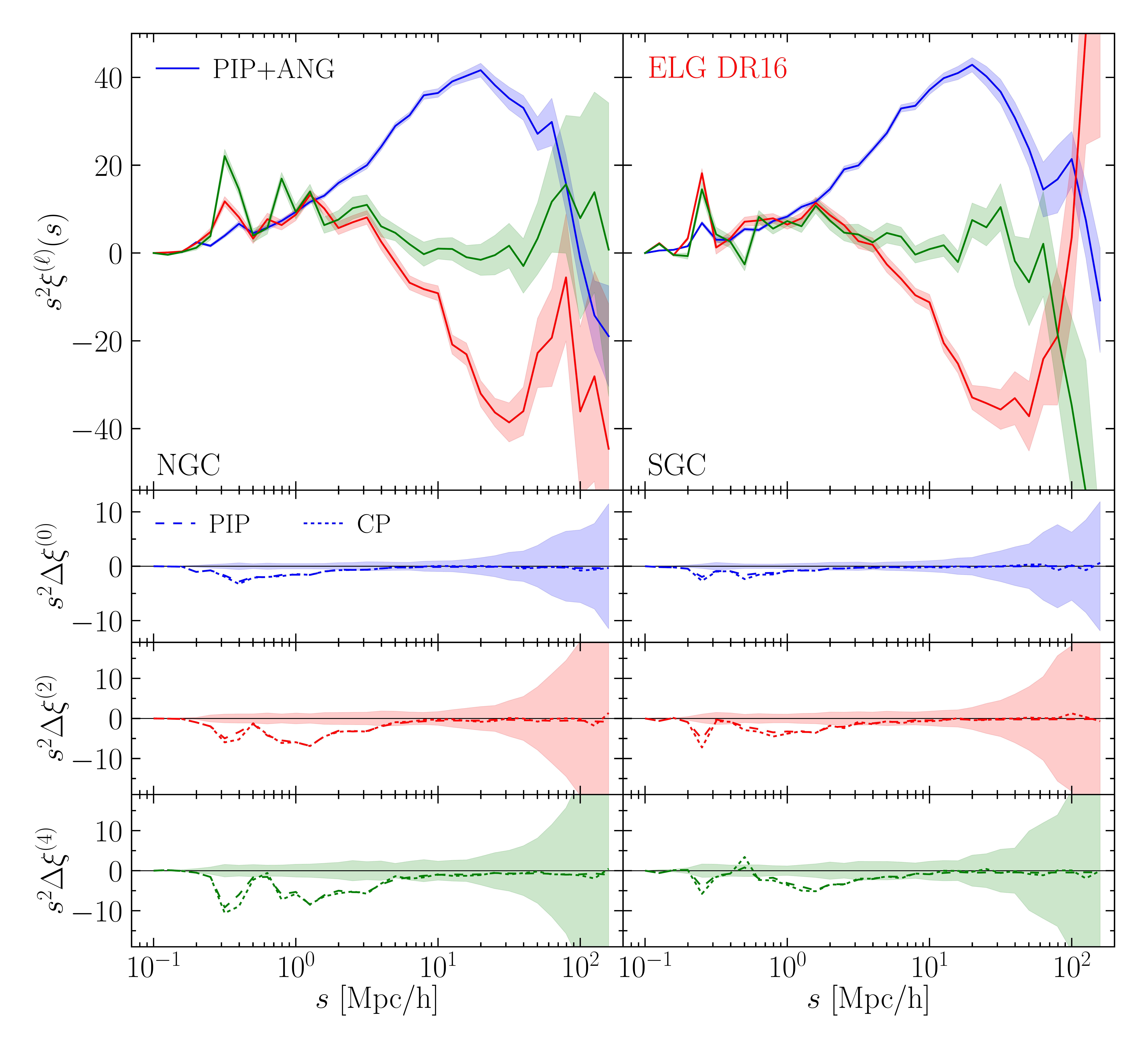}
		\caption{Same as in Fig. \ref{fig:mps_lrg_data}, here for the eBOSS DR16 catalogue of emission-line galaxies.}\label{fig:mps_elg_data}
	\end{figure}

In Fig. \ref{fig:mps_lrg_data}, \ref{fig:mps_qso_data} and \ref{fig:mps_elg_data} we show the measurements of the redshift-space multipole correlation functions for the eBOSS DR16 samples of luminous red galaxies, quasars and emission-line galaxies, respectively. The top panels show only the measurements performed using our reference PIP+ANG up-weighting. In the bottom panels we plot the deviations of the CP and PIP corrections with respect to the measurements that use the joint PIP and angular up-weighting.

As for the projected correlation functions shown in Fig. \ref{fig:wp_lrg_data}-\ref{fig:wp_elg_data}, CP and PIP corrections provide similar results with discrepancies consistent with statistical uncertainties. The difference of the CP and PIP up-weighting with respect to the joint PIP and angular corrections appears to be significant at scales smaller than $\sim10\mhmpc$. Such a difference is due to the presence of zero-probability pairs at transverse separations below the fibre-collision scale as discussed in Sec. \ref{sec:ang_discussion}. These results are consistent with the measurements from the mock catalogues discussed in Sec. \ref{sec:mps_mocks}. At scales below $\sim10\mhmpc$ CP and PIP up-weighting perform better for the quasar catalogue and emission-line galaxies with respect to the sample of luminous red galaxies. This is expected since the luminous red galaxies show a higher small-scale clustering compared to the quasars and emission-line galaxies  (see Fig. \ref{fig:wp_lrg_data}-\ref{fig:wp_elg_data}).

\section{Summary and Conclusions}
We have applied the pairwise-inverse-probability and the angular up-weighting to the 3D clustering measurements from eBOSS DR16 catalogues in order to correct for the systematic bias arising from fibre collisions in the spectroscopic observations. We inferred selection probabilities by means of multiple survey realisations obtained re-running several times the SDSS tiling code on the input target catalogues. We focused on the measurements of the projected correlation function and the multipoles of the redshift-space two-point correlation functions.

We used the approximate ``effective Zel'dovich'' EZmocks to test the performance of the correction method. To this end we edited raw mock catalogues for the luminous-red galaxies, emission-line galaxies and quasars, to obtain synthetic samples as close as possible to real catalogues in terms of number density of targets. These mock catalogues were processed using the same SDSS tiling code used for the actual eBOSS target selection for the spectroscopic follow-up, in order to reproduce fibre collisions. Clustering measurements were then corrected for missing observations using PIP and angular up-weighting schemes. PIP corrections applied to the projected correlation function $\mwp$ provided un-biased measurements on scales larger than $r_p\sim1\mhmpc$ but were strongly biased at smaller scales. This is consistent with the presence of the so called `zero-probability' pairs due to the fibre collisions in single-pass regions. Including the angular up-weighting provided un-biased estimates on scales as small as $0.1\mhmpc$ within data precision. When averaged over 100 EZmocks, a systematic bias was found in the $\mwp$ measurements close to the fiber-collision scales due to the eBOSS strategy of maximising the targeting efficiency in collision groups. These results were reproduced also in the tests on the multipoles of the redshift-space correlation function. However, due to the projection on Legendre polynomials, the systematic bias for the PIP correction alone gets spread over angle-averaged scales as large as $\sim10\mhmpc$. Combining the PIP corrections with angular up-weighting we successfully removed the residual systematic offset in the measured multipoles at scales below $\sim10\mhmpc$. 

The EZmocks used in this work are based on the Zel'dovich approximation. As such they do not faithfully reproduce the small-scale clustering observed in real data. Mock targets exhibit a lower level of clustering compared to the eBOSS DR16 samples (see Fig. \ref{fig:wp_lrg_mocks}-\ref{fig:wp_elg_mocks} and Fig. \ref{fig:wp_lrg_data}-\ref{fig:wp_elg_data}). This makes the fibre-collision issue less severe in the mocks with respect to real data. However, we tested the method using three different tracers. We recovered the input clustering within 1-$\sigma$ errors in all three cases despite the fact that different tracers exhibit significantly different intrinsic clustering strength and are affected by fibre collisions in different ways. This assures us that the technique adopted in this paper is universal and as such does not depend on intrinsic features of a particular sample.

We finally applied the PIP+ANG weighting to the eBOSS DR16 catalogues to correct for fibre-collision when estimating the projected and multipole correlation functions. We compared the joint PIP and angular up-weighting (PIP+ANG) with the PIP-only and with the standard `CP' weighting, a modified version of the nearest-neighbour (NN) method, that is adopted in eBOSS cosmological analyses and previous SDSS spectroscopic samples such as those from the Baryon Oscillation Spectroscopic Survey (BOSS). As expected from tests performed on mock catalogues, the three correction methods perform similarly at scales above $\sim1\mhmpc$ for the projected correlation function and $\sim10\mhmpc$ for the multipole correlation functions, with differences consistent with a statistical noise. On smaller scales both PIP and CP weighting provides a very low clustering compared to the PIP+ANG corrections.

Our analysis provides a robustness test for the standard technique adopted in BOSS and eBOSS cosmological analyses to correct for fibre collisions on scales larger than $~10\mhmpc$. Additionally, the PIP and angular up-weighting scheme tested in this paper is more robust and provides the most accurate 3D clustering measurements down to $\sim\mhkpc$ scales. This gives us access to scales where one-halo term dominates in the clustering signal. As such these scales are crucial to constrain the Halo Occupation Distribution (HOD) models, that study how different tracers populate dark matter haloes of different masses. Moreover, analyses that rely on numerical simulation or HOD formalism to model redshift-space distortions can now be pushed to smaller scales where the clustering signal is measured with high significance. This will allow putting even tighter constraints on the growth rate of structure, a key parameter to constrain gravity models at cosmological scales. These analyses will be presented in future work.\label{sec:conclusions}

\section*{Acknowledgements}
\addcontentsline{toc}{section}{Acknowledgements}

This research was supported by the Centre for the Universe at the Perimeter Institute. Research at Perimeter Institute is supported in part by the Government of Canada through the Department of Innovation, Science and Economic Development Canada and by the Province of Ontario through the Ministry of Economic Development, Job Creation and Trade. We acknowledge support provided by Compute Ontario (www.computeontario.ca) and Compute Canada (www.computecanada.ca).

H.~J.~S. is supported by the U.S.~Department of Energy, Office of Science, Office of High Energy Physics under Award Number DE-SC0014329.

Funding for the Sloan Digital Sky Survey IV has been provided by the Alfred P. Sloan Foundation, the U.S. Department of Energy Office of Science, and the Participating Institutions. SDSS-IV acknowledges
support and resources from the Center for High-Performance Computing at
the University of Utah. The SDSS web site is www.sdss.org.

SDSS-IV is managed by the Astrophysical Research Consortium for the 
Participating Institutions of the SDSS Collaboration including the 
Brazilian Participation Group, the Carnegie Institution for Science, 
Carnegie Mellon University, the Chilean Participation Group, the French Participation Group, Harvard-Smithsonian Center for Astrophysics, 
Instituto de Astrof\'isica de Canarias, The Johns Hopkins University, Kavli Institute for the Physics and Mathematics of the Universe (IPMU) / 
University of Tokyo, the Korean Participation Group, Lawrence Berkeley National Laboratory, 
Leibniz Institut f\"ur Astrophysik Potsdam (AIP),  
Max-Planck-Institut f\"ur Astronomie (MPIA Heidelberg), 
Max-Planck-Institut f\"ur Astrophysik (MPA Garching), 
Max-Planck-Institut f\"ur Extraterrestrische Physik (MPE), 
National Astronomical Observatories of China, New Mexico State University, 
New York University, University of Notre Dame, 
Observat\'ario Nacional / MCTI, The Ohio State University, 
Pennsylvania State University, Shanghai Astronomical Observatory, 
United Kingdom Participation Group,
Universidad Nacional Aut\'onoma de M\'exico, University of Arizona, 
University of Colorado Boulder, University of Oxford, University of Portsmouth, 
University of Utah, University of Virginia, University of Washington, University of Wisconsin, 
Vanderbilt University, and Yale University.

This project has received funding from the European Research Council (ERC) under the European Union's Horizon 2020 research and innovation programme (grant agreement No 693024).





\bibliographystyle{mnras}
\bibliography{biblio_eboss_pip}




\bsp	
\label{lastpage}
\end{document}